\documentclass[aps,11pt,prd,longbibliography]{revtex4-1}

\usepackage{feynmp-auto}
\usepackage{amsmath}
\usepackage[utf8x]{inputenc} 

\usepackage{bbm}
\usepackage{soul}	
\usepackage{xcolor}

\usepackage{epsfig}
\usepackage{color}
\usepackage{slashed}
\usepackage{comment}
\usepackage{epstopdf}

\usepackage{array}
\usepackage{booktabs}
\usepackage{mathrsfs}
\usepackage[euler]{textgreek}
\usepackage{amsmath}
\usepackage{amsthm}
\usepackage{amsfonts}
\usepackage{amssymb}
\usepackage{graphicx}
\graphicspath{ {Figures/} }
\usepackage[font={small}]{caption}
\usepackage{float}
\usepackage{multirow}
\usepackage[export]{adjustbox}
\usepackage[hang, flushmargin,bottom]{footmisc} 
\usepackage{hyperref}
\hypersetup{
     colorlinks   = true,
     citecolor    = blue
}

\usepackage{placeins}
\usepackage{enumitem}

\usepackage{slashed}
\usepackage{bbm}
\usepackage{appendix}

\usepackage{titlesec}
\titleformat{\chapter}[display]
  {\normalfont\LARGE\bfseries}
  {\chaptertitlename\ \thechapter}{5pt}{\LARGE}
  \titlespacing*{\chapter}{0pt}{-20pt}{35pt}
\usepackage{bigstrut}
\usepackage{pst-node}
\usepackage{pstricks}

\usepackage{physics}
\usepackage{mathtools}
\usepackage{setspace}
\usepackage{fancyhdr}
\usepackage[makeroom]{cancel}

\newcommand{\be}{\begin{equation}}
\newcommand{\ee}{\end{equation}}
\newcommand{\bes}{\begin{equation*}}
\newcommand{\ees}{\end{equation*}}

\usepackage{hyperref}% http://ctan.org/pkg/hyperref
\hypersetup{%
  colorlinks = true,
  linkcolor  = black
}
\usepackage{soul}

\newcommand{\beq}{\begin{equation}}
\newcommand{\eeq}{\end{equation}}

\newcommand{\SU}{\,{\rm SU}}
\newcommand{\U}{\,{\rm U}}

\usepackage[normalem]{ulem}

\DeclareUnicodeCharacter{2212}{-}

\AtBeginDocument{%
    \newwrite\bibnotes
    \def\bibnotesext{Notes.bib}
    \immediate\openout\bibnotes=\jobname\bibnotesext
    \immediate\write\bibnotes{@CONTROL{REVTEX41Control}}
    \immediate\write\bibnotes{@CONTROL{%
    apsrev41Control,author="08",editor="1",pages="1",title="0",year="1"}}
     \if@filesw
     \immediate\write\@auxout{\string\citation{apsrev41Control}}%
    \fi
}%

\usepackage{amsfonts}
\usepackage{graphicx}
\usepackage{tcolorbox}
\usepackage{marvosym} 
\usepackage{tikz}
\usepackage{tabulary}
\usepackage{pgfplots}
\usepackage{subfig}
\usepackage{amsmath}
\usepackage{slashed}
\usepackage{mathtools}
\usepackage{nccmath}
\usepackage{verbatim}
\usepackage{color}
\usepackage{amsmath, amsthm, amssymb, amsfonts}
\usepackage{array}
\newcolumntype{P}[1]{>{\centering\arraybackslash}p{#1}}
\usepackage{enumerate}
\usepackage{physics}
\usepackage[font={small, up}]{caption}
\usepackage{tikz}
\usetikzlibrary{"arrows", "automata", "backgrounds", "calendar", "chains", "matrix", "mindmap", "patterns", "petri", "shadows", "shapes.geometric", "shapes.misc", "spy", "trees"}
\graphicspath{{./images/}}
\usepackage{appendix}

\usepackage{tikz}
\usetikzlibrary{arrows,shapes}
\usetikzlibrary{trees}
\usetikzlibrary{matrix,arrows} 				% For commutative diagram
\usetikzlibrary{positioning}				% For "above of=" commands
\usetikzlibrary{calc,through}				% For coordinates
\usetikzlibrary{decorations.pathreplacing}  % For curly braces
\usepackage{pgffor}							% For repeating patterns

\usetikzlibrary{decorations.pathmorphing}	% For Feynman Diagrams
\usetikzlibrary{decorations.markings}
\tikzset{
    vector/.style={decorate, decoration={snake}, draw},
	provector/.style={decorate, decoration={snake,amplitude=2.5pt}, draw},
	antivector/.style={decorate, decoration={snake,amplitude=-2.5pt}, draw},
    fermion/.style={draw=black, postaction={decorate},
        decoration={markings,mark=at position .55 with {\arrow[draw=black]{>}}}},
    fermionuchannel1/.style={draw=black, postaction={decorate},
        decoration={markings,mark=at position .7 with {\arrow[draw=black]{>}}}},
    fermionuchannel2/.style={draw=black, postaction={decorate},
        decoration={markings,mark=at position .4 with {\arrow[draw=black]{>}}}},
    fermionbar/.style={draw=black, postaction={decorate},
        decoration={markings,mark=at position .55 with {\arrow[draw=black]{<}}}},
    fermionnoarrow/.style={draw=black},
    gluon/.style={decorate, draw=black,
        decoration={coil,amplitude=4pt, segment length=5pt}},
    scalar/.style={dashed,draw=black, postaction={decorate},
        decoration={markings,mark=at position .55 with {\arrow[draw=black]{>}}}},
          scalaruchannel1/.style={dashed,draw=black, postaction={decorate},
        decoration={markings,mark=at position .7 with {\arrow[draw=black]{>}}}},
                  scalaruchannel2/.style={dashed,draw=black, postaction={decorate},
        decoration={markings,mark=at position .4 with {\arrow[draw=black]{>}}}},
    scalarbar/.style={dashed,draw=black, postaction={decorate},
        decoration={markings,mark=at position .55 with {\arrow[draw=black]{<}}}},
    scalarnoarrow/.style={dashed,draw=black},
    electron/.style={draw=black, postaction={decorate},
        decoration={markings,mark=at position .55 with {\arrow[draw=black]{>}}}},
	bigvector/.style={decorate, decoration={snake,amplitude=4pt}, draw},
}

\tikzstyle{block} = [draw, rectangle, 
    minimum height=3em, minimum width=6em]

\NewDocumentCommand\semiloop{O{black}mmmO{}O{above}}
{%
\draw[#1] let \p1 = ($(#3)-(#2)$) in (#3) arc (#4:({#4+180}):({0.5*veclen(\x1,\y1)})node[midway, #6] {#5};)
}
\NewDocumentCommand\hello{O{black}mmmO{}O{above}}
{%
\draw[#1] let \p1 = ($(#3)-(#2)$) in (#3) arc (#4:({#4-180}):({0.5*veclen(\x1,\y1)})node[midway, #6] {#5};)
}

\usepackage{hyperref}			% Has to be at the end.
\begin{document}
%%%%%%%%%%%%%%%%%%%%%%%%%%%%%%%%%%%%%%%%%%%%%%%%%%%%%%%%%%%%%%%%%%%%%%%%%%%%%%%%%%

%%%%%%%%%%%%%%%%%%%%%%%%%%%%%%%%%%%%%
\title{\Large {\bf{Baryogenesis via Leptogenesis: \\ Spontaneous $B$ and $L$ Violation}}}
\author{Pavel Fileviez P\'erez$^{1}$, Clara Murgui$^{2}$, Alexis D. Plascencia$^{1}$}
\affiliation{$^{1}$Physics Department and Center for Education and Research in Cosmology and Astrophysics (CERCA), 
Case Western Reserve University, Cleveland, OH 44106, USA \\
$^{2}$Walter Burke Institute for Theoretical Physics, California Institute of Technology, Pasadena, CA 91125}
\email{pxf112@case.edu, cmurgui@caltech.edu, alexis.plascencia@case.edu}
\vspace{1.5cm}

\begin{abstract}
In order to address the baryon asymmetry in the Universe one needs to understand the origin of baryon ($B$) and lepton ($L$) number violation. In this article, we discuss the mechanism of baryogenesis via leptogenesis to explain the matter-antimatter asymmetry in theories with spontaneous breaking of baryon and lepton number. In this context, a lepton asymmetry is generated through the out-of-equilibrium decays of right-handed neutrinos at the high-scale, while local baryon number must be broken below the multi-TeV scale to satisfy the cosmological bounds on the dark matter relic density. We demonstrate how the lepton asymmetry generated via leptogenesis can be converted in two different ways: 
\textit{a)} in the theory predicting Majorana dark matter the lepton asymmetry is converted into a baryon asymmetry, 
and \textit{b)} in the theory with Dirac dark matter the decays of right-handed neutrinos can generate lepton and dark matter asymmetries that are then partially converted into a baryon asymmetry. Consequently, we show how to explain the matter-antimatter asymmetry, the dark matter relic density and neutrino masses in theories for local baryon and lepton number.
\end{abstract}
%%%%%%%%%%%%%%%%
\maketitle 

\hypersetup{linkcolor=blue}

%%%%%%%%%%%%%%%%%%%%%%%%%%%%%%%%%%%%%%%%%%%%%%%%%%%%%%%%%%%%%%%%%%%%%%%%%%%%%%%
\section{INTRODUCTION}
%%%%%%%%%%%%%%%%%%%%%%%%%%%%%%%%%%%%%%%%%%%%%%%%%%%%%%%%%%%%%%%%%%%%%%%%%%%%%%%
The origin of the baryon asymmetry of the Universe remains one of the outstanding open problems in cosmology. The  baryon-to-photon ratio is defined as
\begin{equation}
\eta_B \equiv \frac{n_B-n_{\bar{B}}}{n_\gamma},
\end{equation}
where $n_B$, $n_{\bar{B}}$ and $n_\gamma$ are the number densities of baryons, antibaryons and photons, respectively.
This quantity has been measured from Big-Bang Nucleosynthesis (BBN) \cite{Cooke:2017cwo,Zyla:2020zbs} and 
Cosmic Microwave Background (CMB) radiation data, the current values are \cite{Aghanim:2018eyx}:
\[
\begin{aligned}
{\eta^{\text{BBN}}_{B}}  & = \left(5.80-6.50\right)\times 10^{-10}, \\
{\eta^{\text{CMB}}_{B}}& = \left(6.04-6.20\right)\times 10^{-10},
\end{aligned}
\]
at 95$\%$ CL, respectively. In order to explain the baryon 
asymmetry, the well-known Sakharov conditions have to be satisfied: \textit{a)} Baryon number violation, \textit{b)} C and CP-violation and \textit{c)} the out-of-equilibrium conditions~\cite{Sakharov:1967dj}.
For a review about different baryogenesis mechanisms see Ref.~\cite{Cline:2006ts}.

The origin of the neutrino masses remains one of the open problems in particle physics. One of the simplest mechanisms to generate neutrino masses is the canonical seesaw mechanism~\cite{Minkowski:1977sc,GellMann:1980vs, Mohapatra:1979ia,Yanagida:1979as}. In this scenario, at least two right-handed neutrinos are introduced to generate Majorana masses for the active neutrinos, via the Dirac Yukawa coupling with the Higgs boson in the Standard Model (SM), that are generically suppressed by the mass of the right-handed neutrinos. 

In the context of the canonical seesaw mechanism, there exists a very appealing mechanism to explain the baryon asymmetry in the Universe referred to as Baryogenesis via Leptogenesis~\cite{Fukugita:1986hr}. The main idea is that a lepton asymmetry can be generated through the CP-violating out-of-equilibrium decays of the right-handed neutrinos which is then transferred into a baryon asymmetry 
by the Standard Model sphaleron processes. In this scenario there can be enough CP-violation due to the fact that the Dirac Yukawa couplings can be complex and the non-perturbative baryon number violating sphaleron processes in the SM can be used. For more details we refer to the reviews in Refs.~\cite{Buchmuller:2004nz,Buchmuller:2005eh,Davidson:2008bu,Fong:2013wr}.

\begin{figure}[b]
\centering
\includegraphics[width=0.95\linewidth]{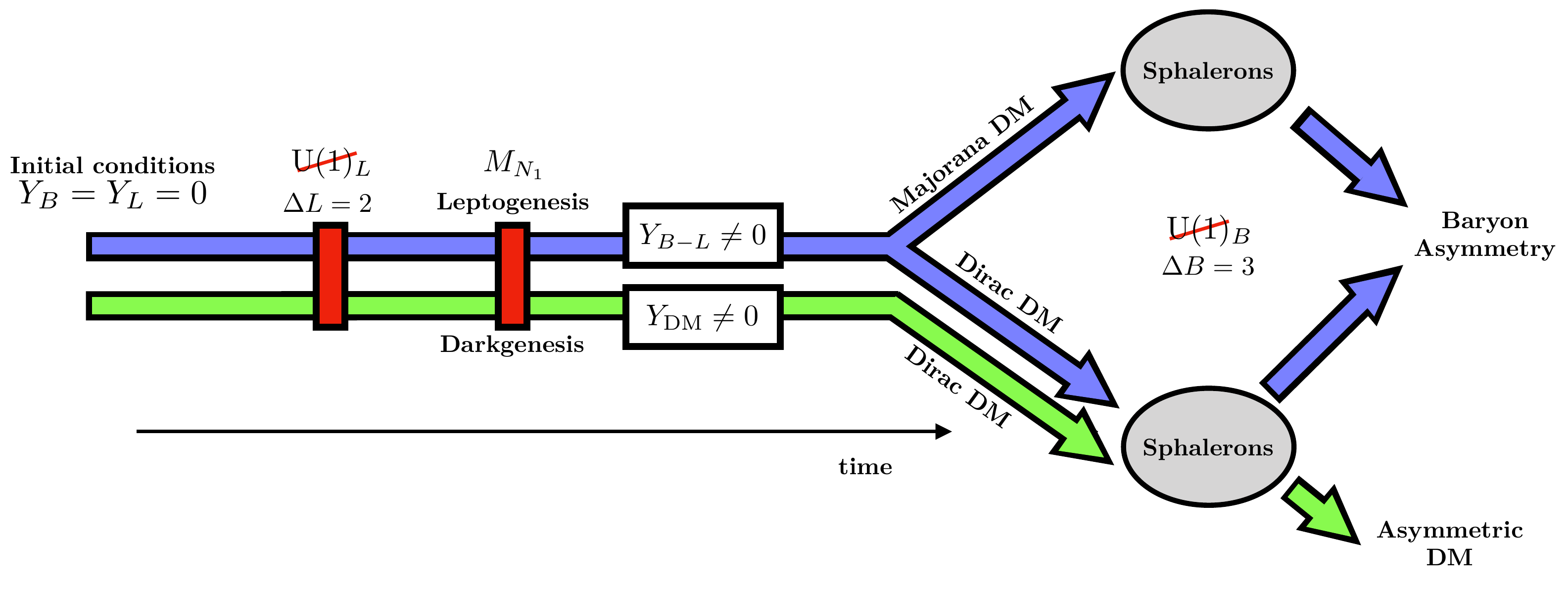}
\caption{Different mechanisms for baryogenesis in theories for spontaneous breaking of local $B$ and $L$.}
\label{Mechanisms}
\end{figure}

The nature of the dark matter in the Universe constitutes another one of the open problems in cosmology. There are many appealing candidates that can explain the dark matter relic density in the Universe such as axions or weakly interacting massive particles (WIMPs). An alternative explanation is having an asymmetric population of dark matter, and then, the baryon asymmetry and the dark matter relic density can be related, see Refs.~\cite{Petraki:2013wwa,Zurek:2013wia} for reviews on this topic.

In this article we study the possibility to implement the mechanism of Baryogenesis via Leptogenesis in theories 
where Baryon and Lepton numbers are local gauge symmetries~\cite{Perez:2014qfa,Duerr:2013dza}. 
In this context, before spontaneous symmetry breaking occurs there cannot be any initial lepton or baryon asymmetries. 
These theories have two additional scales, the $\U(1)_B$ and the $\U(1)_L$ breaking scales.
Moreover, these theories predict a dark matter candidate from anomaly cancellation and there exists an upper bound on the $\U(1)_B$ scale 
below the multi-TeV scale after imposing the dark matter relic density constraints~\cite{FileviezPerez:2019jju,FileviezPerez:2018jmr}.
Three right-handed neutrinos are needed to cancel the leptonic gauge anomalies and their masses are generated 
when $\U(1)_L$ is broken. One can say that the upper bound on the $\U(1)_L$ breaking scale is basically the well-known upper bound on 
the seesaw scale, $10^{15-16}$ GeV. In this article, we demonstrate that there is a conservative lower bound on the $\U(1)_L$ 
scale assuming that the baryon asymmetry in the Universe can be explained through leptogenesis.

The realization of the leptogenesis mechanism in theories with local $\U(1)_L$ is different from the case with the canonical seesaw mechanism. In the former, the right-handed neutrinos 
have new interactions with the gauge boson associated to $\U(1)_L$ which play an important role. These interactions mediated by the new gauge boson, $Z_L$, 
can keep the right-handed neutrinos in equilibrium and will have an impact on the evolution for the particle abundances around the time of leptogenesis. 
On the one hand, this new interaction thermalizes the right-handed neutrinos in early times so that an equilibration particle abundance can be easily achieved 
before leptogenesis. On the other hand, if the interaction is too large then the right-handed neutrinos will not be driven out-of-equilibrium. In consequence, 
by requiring leptogenesis to explain the measured baryon asymmetry we will demonstrate there is a conservative lower bound on the symmetry breaking scale for lepton number; namely, 
$M_{Z_L}/g_L \gtrsim 10^{10}$ GeV. Previous studies have focused on studying leptogenesis in theories with $B-L$ as a local symmetry~\cite{Enqvist:1992nf,Plumacher:1996kc,Racker:2008hp,Blanchet:2009bu,Iso:2010mv,Heeck:2016oda,Dev:2017xry,Bian:2019szo} or on the effect of a high-scale phase transition on the $N_1$ departure from equilibrium~\cite{Shuve:2017jgj}. 

When we study the mechanism of leptogenesis in theories for spontaneous breaking of local $B$ and $L$ there exists a strong link to the nature of the dark matter candidate. There are two simple realizations of these theories, see Refs.~\cite{Perez:2014qfa,Duerr:2013dza} for details. 
In the first of these theories~\cite{Perez:2014qfa}, the dark matter candidate is a Majorana fermion and the baryon 
asymmetry is generated after the lepton asymmetry is converted by the sphalerons even when $\U(1)_B$ is broken at the 
electroweak scale. In the second class of theories~\cite{Duerr:2013dza}, the dark matter can be a Dirac fermion and in general there can be an asymmetry in the dark sector. Therefore, in these theories there can be lepton and dark matter asymmetries generated
through the out-of-equilibrium decays of the right-handed neutrinos. These asymmetries then get redistributed to a baryon asymmetry by sphaleron processes even when $\U(1)_B$ is broken at scales close to the electroweak scale. In Fig.~\ref{Mechanisms} we show a schematic diagram of the different mechanisms for baryogenesis in theories for spontaneous $B$ and $L$ breaking.

This work is structured as follows. In Section~\ref{sec:leptogenesis} we give an overview of the mechanism of leptogenesis, discuss the implications of the  new scattering processes and present our main results. In Section~\ref{sec:models} we discuss different mechanisms for baryogenesis in theories with local baryon and lepton numbers. In Section~\ref{sec:DM} we discuss how a dark matter asymmetry can also be generated from the decays of the right-handed neutrino. The final baryon asymmetry is given in terms of the $B-L$ and the dark matter asymmetries. We summarize our main findings in Section~\ref{sec:summary}.
%
%%%%%%%%%%%%%%%%%%%%%%%%%%%%%%%%%
\section{LEPTOGENESIS}
\label{sec:leptogenesis}
%%%%%%%%%%%%%%%%%%%%%%%%%%%%%%%%%
In this section we discuss how the mechanism of leptogenesis through out-of-equilibrium decays of the right-handed neutrinos can be implemented in theories with local lepton number. In these theories there are new gauge interactions mediated by the $\U(1)_L$ gauge boson, $Z_L$, and our main goal is to find a conservative lower bound 
on the $\U(1)_L$ scale using the out-of-equilibrium condition for the right-handed neutrinos and the observed value of the baryon asymmetry. In order to simplify our analysis, 
we focus on the scenarios where $M_{N_i} \ll M_{Z_L}, M_{S_L}$, and hence, the $\U(1)_L$ gauge boson and the new Higgs $S_L$ responsible for the symmetry breaking can be integrated out. The scalar $S_L$ is a singlet under the SM gauge group and carries two units of lepton number. For studies of leptogenesis in the context of lepton number as a global symmetry see e.g.~\cite{Pilaftsis:2008qt,Alonso:2020com}.

\begin{figure}[t]
\centering
\begin{equation*}
\begin{gathered}
\resizebox{.25\textwidth}{!}{%
\begin{tikzpicture}[line width=1.5 pt,node distance=1 cm and 1.5 cm]
\coordinate[label = left: $N_1$] (p1);
\coordinate[right = of p1](p2);
\coordinate[above right = of p2, label=right: $ H$] (v1);
\coordinate[below right = of p2,label=right: $ \ell_L$] (v2);
\draw[fermionnoarrow] (p1) -- (p2);
\draw[scalar] (p2) -- (v1);
\draw[fermion] (p2) -- (v2);
\draw[fill=red] (p2) circle (.1cm);
\end{tikzpicture}}
\end{gathered}
\!\!\! + \quad 
\begin{gathered}
\resizebox{.32\textwidth}{!}{%
\begin{tikzpicture}[line width=1.5 pt,node distance=1 cm and 1.5 cm]
\coordinate[label = left: $N_1$] (p1);
\coordinate[right = 0.8cm  of p1](p1a);
\coordinate[right = of p1a](p1b);
\coordinate[right = 0.9 cm of p1b](p2);
\coordinate[above right = of p2, label=right: $ H$] (v1);
\coordinate[below right = of p2,label=right: $ \ell_L$] (v2);
\coordinate[above right = 1cm of p1a,label=$\ell_L$](vaux1);
\coordinate[below right = 1cm of p1a,label=$H$](vaux2);
\coordinate[right=0.5cm of p1b,label=above:$N_{j\neq1}$](vaux3);
\draw[fermionnoarrow] (p1) -- (p1a);
\draw[scalar] (p2) -- (v1);
\draw[fermion] (p2) -- (v2);
\draw[fermionnoarrow] (p1b)--(p2);
\semiloop[fermionnoarrow]{p1a}{p1b}{0};
\hello[scalarnoarrow]{p1a}{p1b}{0};
\draw[fill=red] (p2) circle (.1cm);
\draw[fill=red] (p1a) circle (.1cm);
\draw[fill=red] (p1b) circle (.1cm);
\end{tikzpicture}}
\end{gathered}  \!\!\! + \quad 
\begin{gathered}
\resizebox{.3\textwidth}{!}{%
\begin{tikzpicture}[line width=1.5 pt,node distance=1 cm and 1.5 cm]
\coordinate[label = left: $N_1$] (p1);
\coordinate[right = of p1](p2);
\coordinate[right=0.55 cm of p2](vmare);
\coordinate[above=0.5cm of vmare,label=$\ell_L$](vaux1);
\coordinate[below=1.1cm of vmare,label=$H$](vaux2);
\coordinate[right= 1cm of vmare,label=right:$N_{j\neq1}$](vaux3);
\coordinate[above right = of p2](v1a);
\coordinate[right = of v1a, label=right: $H$] (v1);
\coordinate[below right = of p2](v2a);
\coordinate[right = of v2a,label=right: $ \ell_L$] (v2);
\draw[fermionnoarrow] (p1) -- (p2);
\draw[scalar] (v1a) -- (v1);
\draw[fermionnoarrow] (v1a)--(v2a);
\draw[fermion](v1a)--(p2);
\draw[scalar] (v2a)--(p2);
\draw[fermion] (v2a) -- (v2);
\draw[fill=red] (p2) circle (.1cm);
\draw[fill=red] (v1a) circle (.1cm);
\draw[fill=red] (v2a) circle (.1cm);
\end{tikzpicture}}
\end{gathered}
\end{equation*}
\caption{Feynman diagrams for the decays of the lightest right-handed neutrino $N_1$.}
\label{fig:loopdiagrams}
\end{figure}
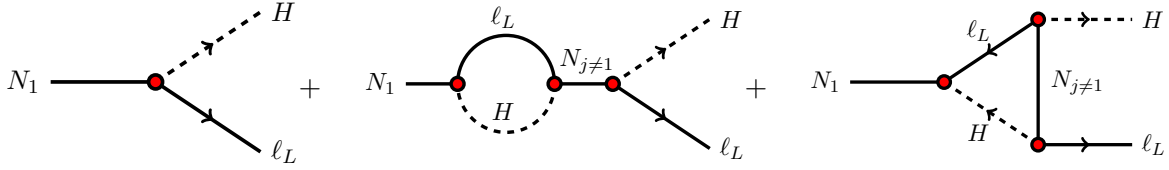
The relevant interactions for our study are given by
\begin{eqnarray}
- \mathcal{L}_\nu &\supset&  g_L \overline{\nu}_R \gamma_\mu Z_L^\mu \nu_R + \left( Y^D_{\alpha i} \, \overline{\ell^\alpha_L} \widetilde{H} \nu_{R}^i + y_R \, \nu_R \nu_R S_L^* + \textrm{h.c.} \right),
\label{eq:YukawaLepto}
\end{eqnarray}
where $\widetilde{H}=i \sigma_2 H^*$ is the SM Higgs boson with SM quantum numbers $H \sim (\mathbf{1}, \mathbf{2}, \tfrac{1}{2})$. In this model the right-handed neutrinos have an extra interaction with the gauge boson associated to local lepton number, $Z_L$. Assuming a hierarchical spectrum for the right-handed neutrinos, the main contribution to the lepton asymmetry comes from the out-of-equilibrium decay of the lightest right-handed neutrino. 

The relevant Boltzmann equations for the evolution of the particle abundances $Y_i = n_i/s$ (where $n_i$ are the number densities and $s$ is the entropy density of the Universe) for the lightest right-handed neutrino and the lepton asymmetry are given by
\begin{align}
\label{eq:Boltzmann1}
\frac{ dY_{N_1}} {dz} = &  - \frac{z}{ s H(M_{N_1})} \left[ \left( \frac{Y_{N_1}}{Y^{\rm eq}_{N_1}} - 1 \right) \gamma_N  + 
\left( \left( \frac{Y_{N_1}}{Y^{\rm eq}_{N_1}} \right)^2 -1 \right) \gamma_{NN}  \right], \\
\label{eq:Boltzmann2}
\frac{ dY_{B-L}} {dz} = & - \frac{z}{ s H(M_{N_1})} \left[ \left( \frac{1}{2} \frac{Y_{B-L}}{Y_\ell^{\rm eq}} - \varepsilon_1  \left(\frac{Y_{N_1}}{Y^{\rm eq}_{N_1}} - 1 \right) \right) \gamma_{D} + \right. 
\left. \frac{Y_{B-L}}{Y^{\rm eq}_\ell} \gamma_{B-L} \right],
\end{align}
where 
\begin{align}
\gamma_N &=\gamma_{D} + 2\gamma_4 + 4\gamma_6 + 2 \gamma_9 + 2\gamma_{10} + 2 \gamma_{11}, \\
\gamma_{NN} &= \gamma_1 + \gamma_2 + \gamma_3 + \gamma_5, \\ 
\gamma_{B-L} &= 2 \gamma_6 + 2 \gamma_7 + 2 \gamma_8 + \gamma_9 + \gamma_{11} + \frac{Y_{N_1}}{Y^{\rm eq}_{N_1}} \gamma_4 + \frac{Y_{N_1}}{Y^{\rm eq}_{N_1}} \gamma_{10}.
\end{align}
In the above equations $z= M_{N_1}/T$ and at equilibrium we have $Y^{\rm eq}_a  = 45 g_a z_a^2 K_2(z_a)/(4 \pi^4 g_*)$, where due to the new states in the theory the total number of effective degrees of freedom is $g_*=116.875$. $K_2(z)$ corresponds to the modified Bessel functions of the second kind. Here the $\gamma$-quantities have the information of the different decay 
and scattering processes producing a lepton asymmetry or changing the $N_1$ particle abundance, 
see the Appendix for details. Notice that only $\gamma_1$ and $\gamma_2$ contain gauge interactions mediated by the $\U(1)_L$ gauge boson. 

The processes mediated by the Dirac Yukawa coupling are expected to be subleading in comparison to the new gauge interactions (whose strength we assume to be order one). 
Since we are interested in finding the lower bound on the mass of the new gauge boson, we focus on the regime 
where the right-handed neutrinos are much below the canonical seesaw scale, where the Yukawa couplings are small.
Therefore, we neglect the scattering processes mediated by Yukawa couplings and the relevant Boltzmann equations are given by
\begin{align}
\label{eq:Boltzmann3}
\frac{ dY_{N_1}} {dz} \simeq &  - \frac{z}{ s H(M_{N_1})} \left[ \left( \frac{Y_{N_1}}{Y^{\rm eq}_{N_1}} - 1 \right)  \gamma_{D}    + \left( \left( \frac{Y_{N_1}}{Y^{\rm eq}_{N_1}} \right)^2 -1 \right) \left(  \gamma_1 + \gamma_2  \right)  \right], \\[2ex]
\label{eq:Boltzmann4}
\frac{ dY_{B-L}} {dz} \simeq & - \frac{z}{ s H(M_{N_1})} \left[ \left( \frac{1}{2} \frac{Y_{B-L}}{Y_\ell^{\rm eq}} - \varepsilon_1  \left(\frac{Y_{N_1}}{Y^{\rm eq}_{N_1}} - 1 \right) \right) \gamma_{D}  \right]. 
\end{align}
Some of the $\gamma$ terms we have dropped have the same dependence as $\gamma_D$; however, the decay term dominates over scattering terms.
Generically, the final lepton asymmetry will depend on how it is distributed among the three flavors. However for simplicity we ignore flavor effects and solve the Boltzmann equation in the one-flavor regime given above.
\begin{figure}[t]
\centering
\hspace{1cm} $M_{N_1}=10^{10}$ GeV \,\,\,\,\,\, $M_{Z_L} = 10^{13}$  GeV \,\,\,\,\,\, $g_L=1.0$ \,\,\,\,\,\, $\varepsilon_1 = 6 \cdot 10^{-8}$\,\,\,\,\,\,\,\,$\widetilde{m}_1 = 10^{-4}$ eV\\
\includegraphics[width=0.495\linewidth]{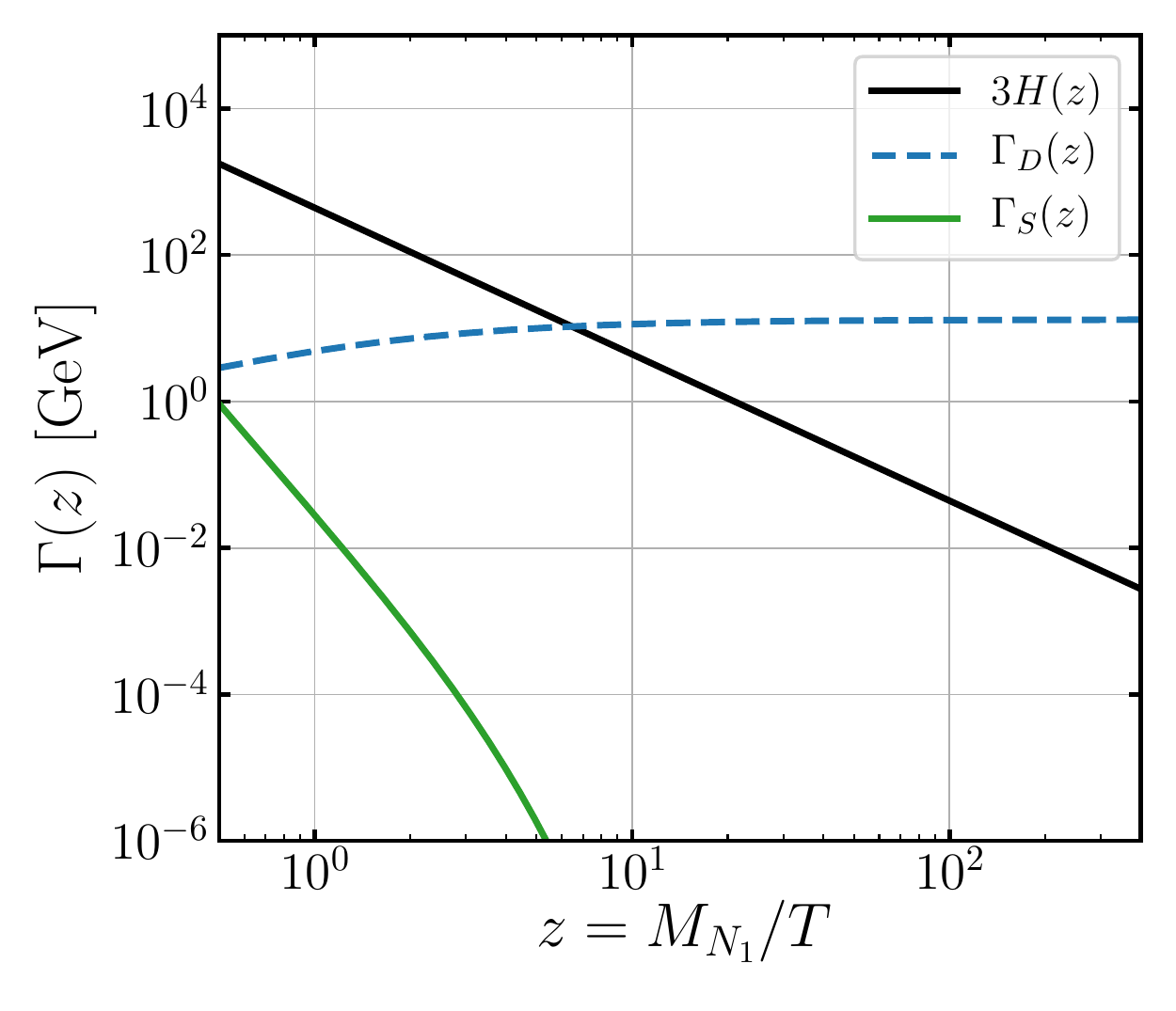}
\includegraphics[width=0.495\linewidth]{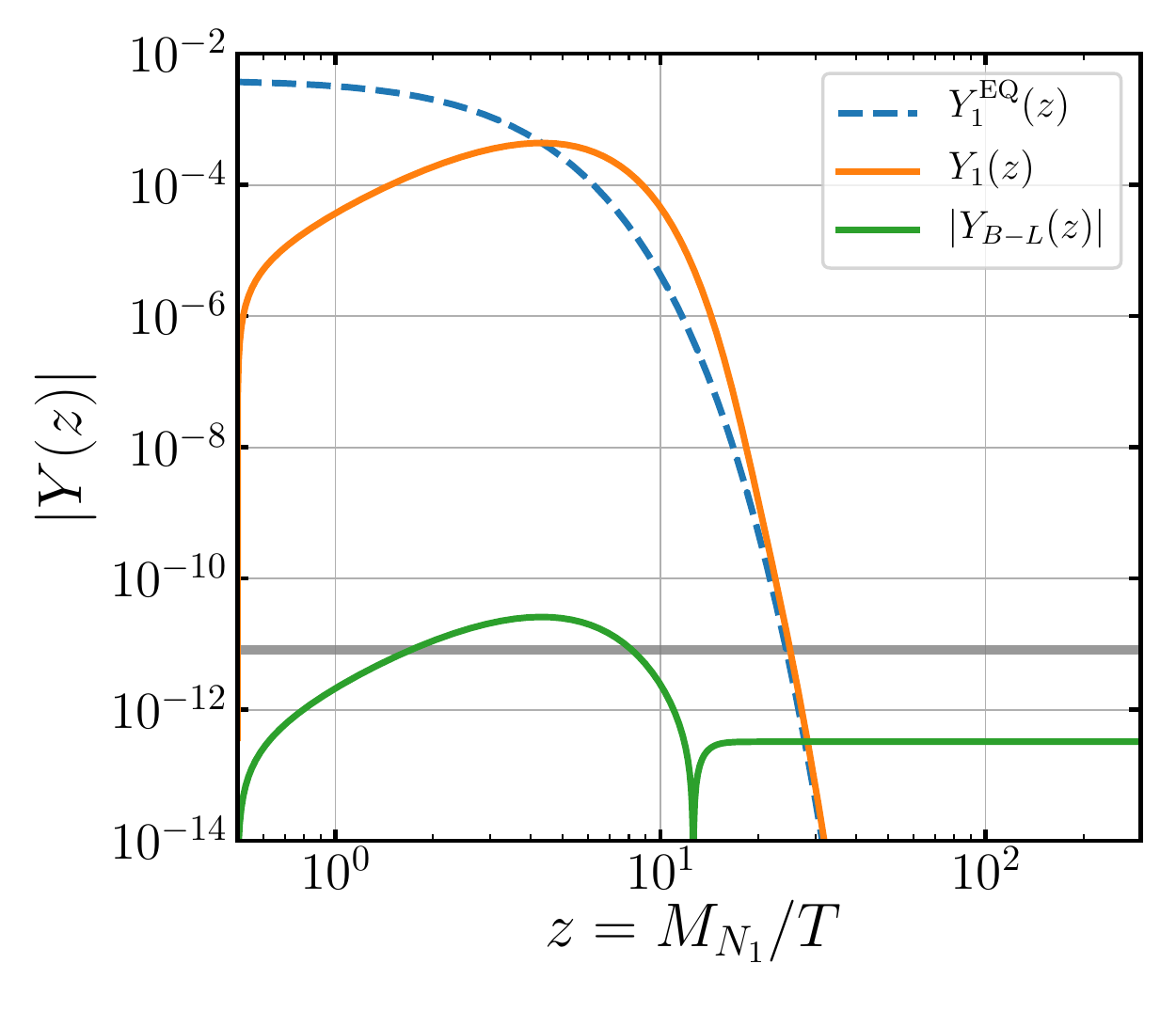}
\phantom{.}\hspace{1cm}  $M_{N_1}=10^{10}$ GeV \,\,\,\,\,\, $M_{Z_L} = 10^{12}$  GeV \,\,\,\,\,\, $g_L=1.0$ \,\,\,\,\,\, $\varepsilon_1 = 6 \cdot 10^{-8}$\,\,\,\,\,\,\,\,$\widetilde{m}_1 = 10^{-3}$ eV\\
\includegraphics[width=0.495\linewidth]{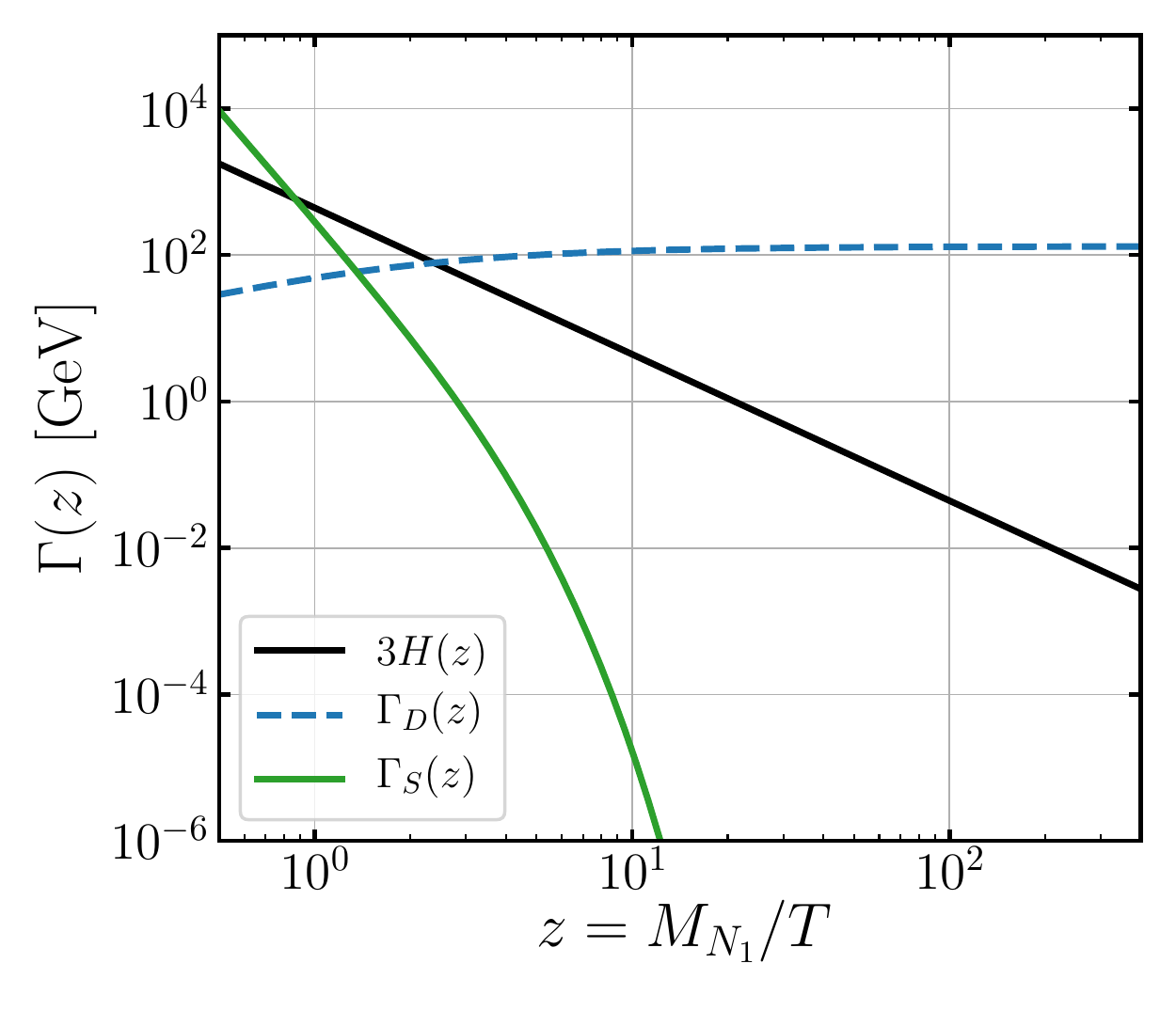}
\includegraphics[width=0.495\linewidth]{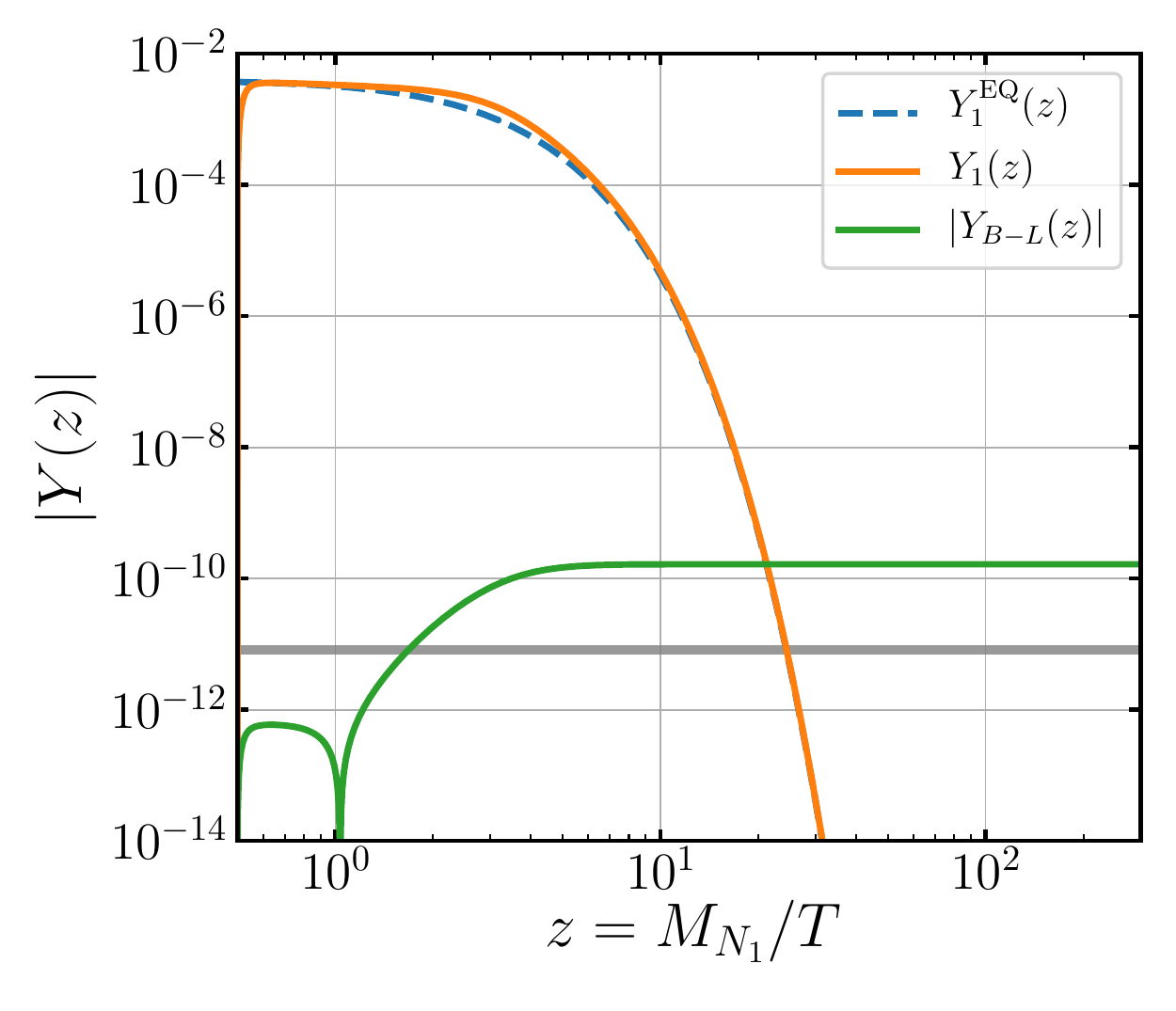}
\caption{ \textit{Left panel:} Different interaction rates as a function of $z= M_{N_1}/T$. The blue dashed line corresponds to the decay rate of $N_1$, while the green solid line corresponds to the scattering rate for $N_1 N_1 \to Z_L^* \to f \bar{f}$. The Hubble expansion rate is shown by the black solid line. \textit{Right panel:} Evolution of the particle abundances as a function of the parameter $z$. The dashed blue line gives the equilibrium abundance for $N_1$, the orange (green) line gives the abundance for $N_1$ (the $B-L$ asymmetry) as solutions to the Boltzmann equations. The gray band shows the $B-L$ asymmetry that corresponds to the observed baryon asymmetry at later times, as given by Eq.~\eqref{eq:211}.}
\label{fig:rates}
\end{figure}

The lepton CP asymmetry is generated from the interference between the tree-level and one-loop contributions to the $N_1$ decays shown in Fig.~\ref{fig:loopdiagrams} (see e.g~\cite{Davidson:2008bu,Blanchet:2012bk})
 \beq
 \varepsilon_1 = \frac{1}{8 \pi}\frac{1}{(Y_D^\dagger Y_D)_{11}}  \sum_{j \neq 1} {\rm Im} \left[ (Y_D^\dagger Y_D)^2_{1j}\right] g \left( \frac{M_{N_j}^2}{M_{N_1}^2} \right),
 \label{eq:epsilon}
 \eeq
where the loop function reads as
\beq
g(x) = \sqrt{x} \left(  \frac{1}{1-x} + 1 - (1+x)\ln \left( \frac{1+x}{x} \right) \right).
\label{eq:gloop}
\eeq
In this theory, $\eta_B$ is related to the particle abundance by 
\beq
\label{eq:211}
\eta_B \approx 211 C\, Y_{B-L}(T_{\rm lepto}),
\eeq
 where $T_{\rm lepto}$ is the temperature at which leptogenesis takes place and $C$  corresponds to  the conversion factor $Y_{B}= C \, Y_{B-L}$ where $C=32/99$ if the phase transition for $\U(1)_B$ occurs simultaneously with the electroweak phase transition~\cite{Perez:2014qfa}; if the new states have already decoupled by the time of the electroweak phase transition then we use the SM value of $C=28/79$~\cite{Harvey:1990qw}.

\begin{figure}[h]
\centering
$M_{N_1}=10^9$ GeV \,\,\,\,\, $M_{Z_L} = 25 \,M_{N_1}$ \,\,\,\,\, $g_L=1.0$ \,\,\,\,\, $\varepsilon_1 = \varepsilon_1^{\rm DI} = 9.9 \cdot 10^{-8}$\\
\includegraphics[width=0.49\linewidth]{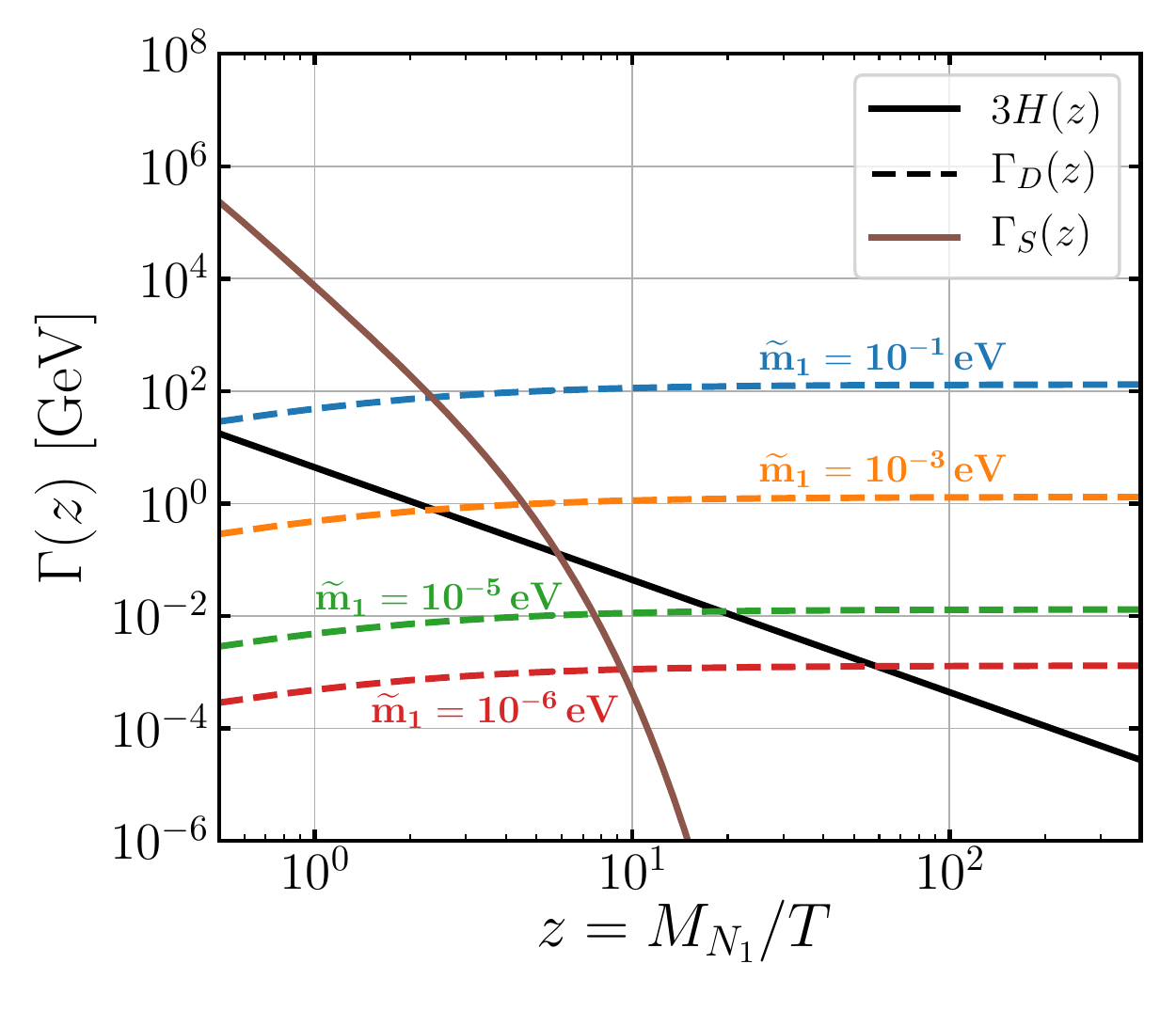}
\includegraphics[width=0.49\linewidth]{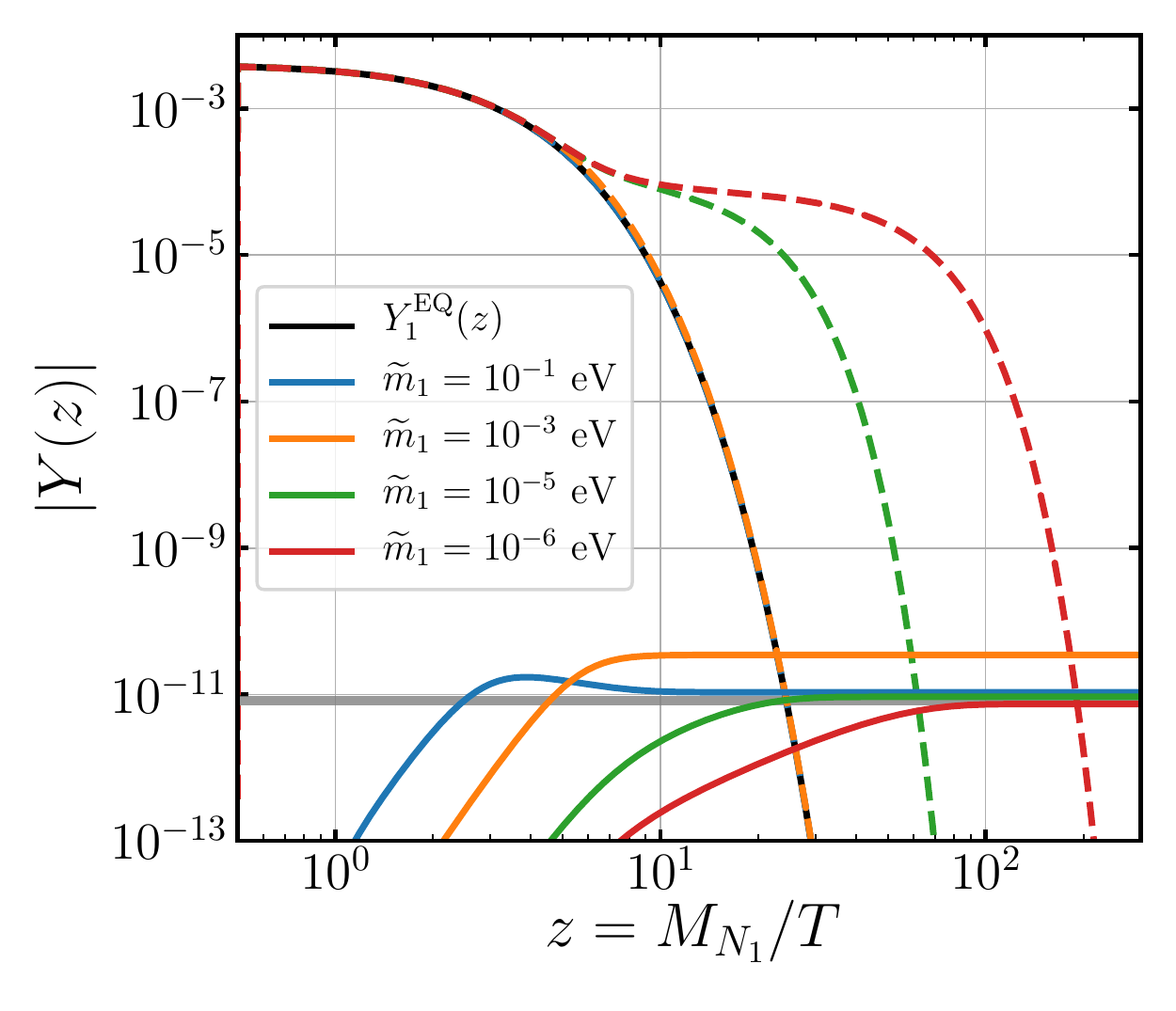}
$M_{N_1}=10^8$ GeV \,\,\,\,\, $M_{Z_L} = 25 \,M_{N_1}$ \,\,\,\,\, $g_L=1.0$ \,\,\,\,\, $\varepsilon_1 = \varepsilon_1^{\rm DI} = 9.9 \cdot 10^{-9}$\\
\includegraphics[width=0.49\linewidth]{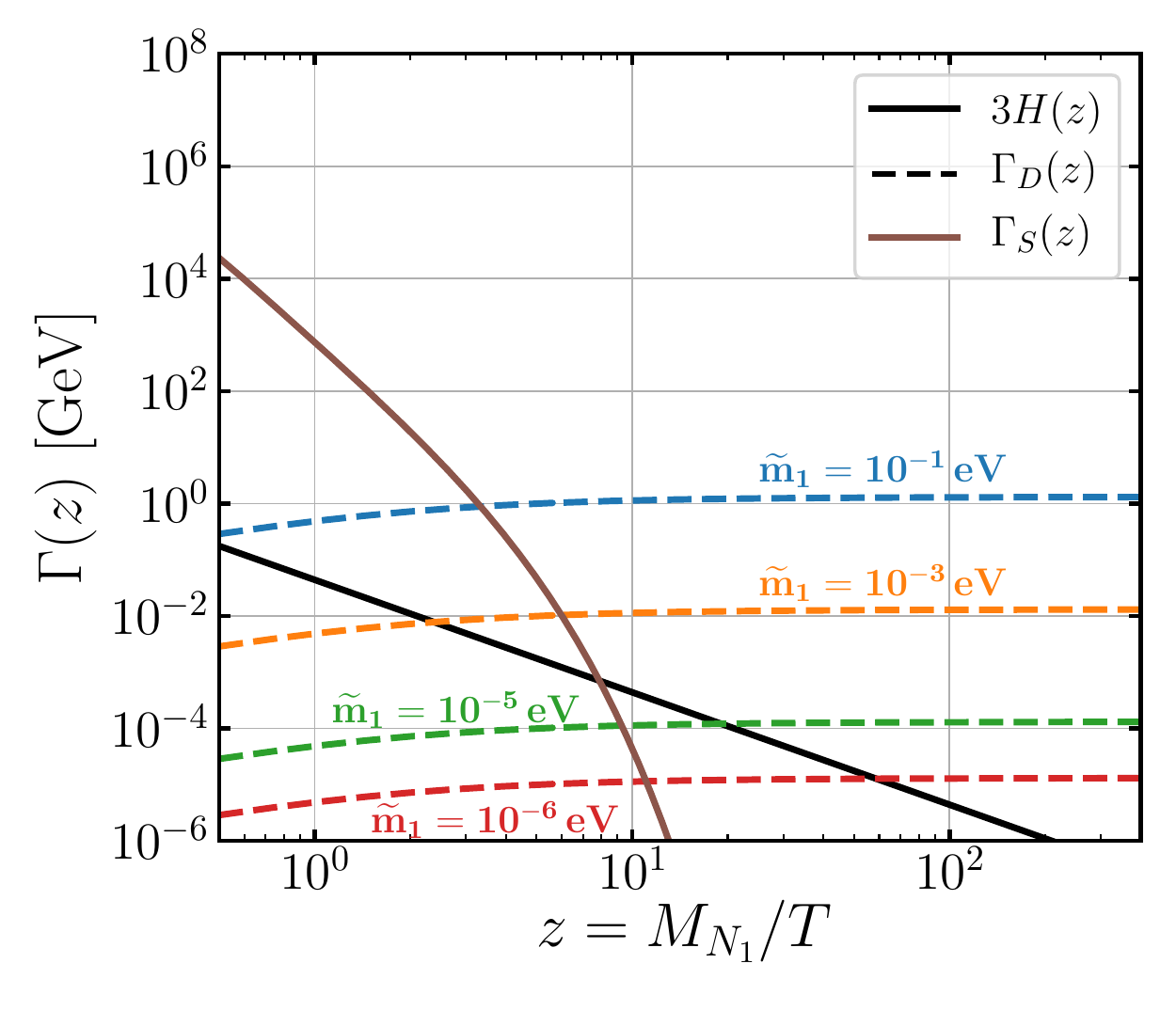}
\includegraphics[width=0.49\linewidth]{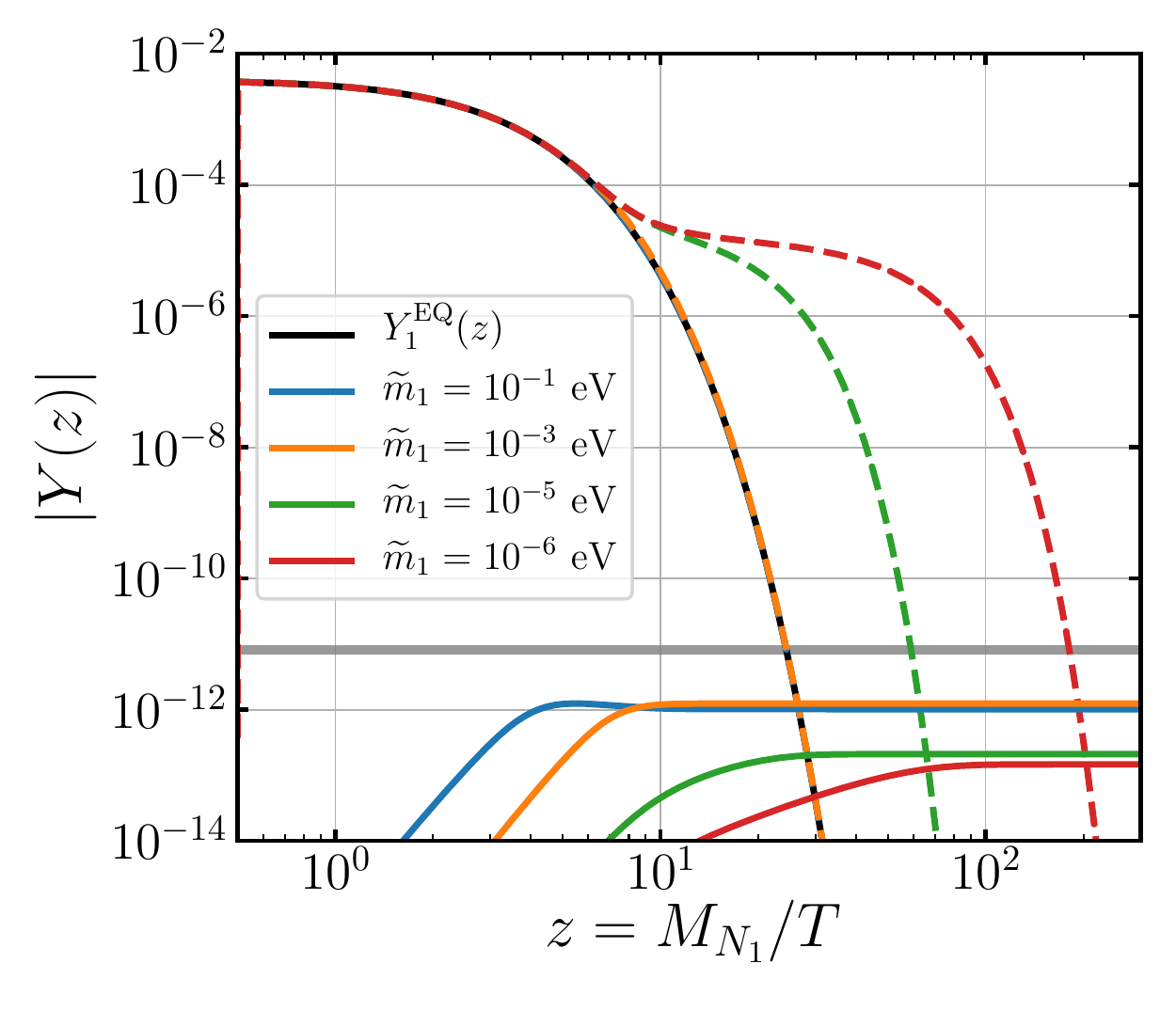}
\caption{Same as Fig.~\ref{fig:rates}. Here we fix $\varepsilon_1$ to its maximal value given in Eq.~\eqref{eq:epsilonDI} and show the results for different values of the decay parameter $\widetilde{m}_1$ as described in the plot legends.}
\label{fig:m1tilde}
\end{figure}

The decay parameter in the Boltzmann equations is given by
\beq
\label{eq:gammaD}
\gamma_D = n^{\rm eq}_{N_1} (z) \frac{K_1(z)}{K_2(z)} \widetilde{\Gamma}_D,
\eeq
where $K_1(z)$ and $K_2(z)$ correspond to the modified Bessel functions of the second kind, and the decay rate is given by
\beq
\label{eq:DecayN}
\widetilde{\Gamma}_D = \widetilde{\Gamma}(N_1 \to \ell_L H) + \widetilde{\Gamma}(N_1 \to \bar{\ell}_L H^\dagger) = \frac{1}{8 \pi} (Y_D^\dagger Y_D)_{11} M_{N_1} =\frac{ M_{N_1}^2}{4\pi v^2} \, \widetilde{m}_1 ,
\eeq
where the last relation gives the definition of the $\widetilde{m}_1$ parameter.
The $\gamma$ parameter for a generic $ab \leftrightarrow cd$ scattering process is given by
\beq
\label{eq:gammafinal}
\gamma(ab  \leftrightarrow cd) = \frac{g_a g_b T}{32 \pi^4} \int_{(m_a+m_b)^2}^\infty ds \, \sigma(ab \to cd) \frac{\lambda(s,m_a^2,m_b^2)}{\sqrt{s}} K_1 \left( \frac{\sqrt{s}}{T} \right),
\eeq
where $\lambda(a, b, c)= (a-b-c)^2 - 4bc \,$ is the K{\"a}ll{\'e}n function and $\sigma(ab \to cd)$ corresponds to the standard cross-section. For $N N \to Z_L^* \to f \overline{f}$ where the final states are $f=\ell, \,F$ (where $\ell$ corresponds to all leptons in the SM and $F$ to the new anomaly-canceling fermions) the cross-section is given by,
\beq
\sigma(N N \to f \overline{f}) = \frac{C_f g_L^4 \sqrt{s} }{8 \pi } \frac{\sqrt{s-4M_N^2} }{[(s-M_{Z_L}^2)^2 + \Gamma_{Z_L}^2 M_{Z_L}^2]} \approx  \frac{C_f g_L^4 \sqrt{s} }{8 \pi } \frac{\sqrt{s-4M_N^2} }{\left( M_{Z_L}^4 + \Gamma_{Z_L}^2 M_{Z_L}^2 \right)} ,
\eeq
where $C_\ell=3$ for the sum over leptons and $C_F=6$ for the sum over the anomaly-canceling fermions. Furthermore, since we are evaluating at very high temperatures we ignore the masses of the final states. The decay width of the gauge boson is given by
\beq
\Gamma_{Z_L} = \frac{9 g_L^2 M_{Z_L}}{8 \pi } + \frac{g_L^2}{12 \pi M_{Z_L}^2} \sum_{i=1}^3 \left( M_{Z_L}^2 - 4M_{N_i}^2\right)^{3/2}.
\eeq

In order to get an idea of when the decay and scattering rates come into thermal equilibrium we can compare them with the Hubble expansion rate $3H(z)$. The decay rate is given by
\beq
\Gamma_D(z) = \frac{K_1(z)}{K_2(z)} \widetilde{\Gamma}_D = \frac{\gamma_D}{n_{N_1}^{\rm eq}(z)},
\eeq
and the equilibration mass is found by setting $\Gamma_D = H$ at $T=M_{N_1}$ or $z=1$, giving
\beq
\widetilde{m}^* = 1.12 \times 10^{-3} \,\, {\rm eV}.
\eeq
The interaction rate for the scattering $N_1 N_1 \to Z_L^* \to f \bar{f}$ is given by
\beq
\Gamma_S (z) = n_{N_1}^{\rm eq}(z) \langle \sigma v \rangle = \frac{( \gamma_1 + \gamma_2 )}{n_{N_1}^{\rm eq}(z)}.
\eeq
The left panels in Fig.~\ref{fig:rates} show the different interaction rates as a function of $z$. 
In order to quantify whether the interaction rate is in equilibrium in the plasma we compare it to the Hubble expansion rate $3H(z)$. 
In the first scenario we set $M_{Z_L}\!=\!10^3M_{N_1}\!=\!10^{13}$ GeV and, as the upper left panel in Fig.~\ref{fig:rates} shows, the scattering cross-section mediated by the $Z_L$ does not thermalize. The later can also be visualized in the right panel where it can be seen that the abundance of $N_1$ does not reach thermal equilibrium at early times. However, as we lower the value of $M_{Z_L}$ we can see that the interaction thermalizes at different values of $z$ which are relevant for leptogenesis, as it is reflected in the lower panels. 

The right panels in Fig.~\ref{fig:rates} show the evolution of the particle abundance for $N_1$ and for the $B-L$ asymmetry. For all our numerical results we set an initial zero abundance for the right-handed neutrino, i.e. $Y_{N_1}(z_0\!=\!0.5)=0$, and as can be seen from our results whenever $M_{Z_L}$ is close to $M_{N_1}$ the process mediated by $Z_L$ very quickly brings $N_1$ into thermal equilibrium.

Assuming a hierarchical spectrum of right-handed neutrinos, it was shown in Ref.~\cite{Davidson:2002qv} that there is an upper bound on the CP asymmetry that only depends on $M_{N_1}$
\beq
\label{eq:epsilonDI}
\varepsilon_1^{\rm max}  (M_{N_1})= \frac{3 M_{N_1}}{8 \pi v^2} (m_3-m_1),
\eeq
where $v=246$ GeV is the SM Higgs vacuum expectation value, and $m_1$ and $m_3$ correspond to the lightest and heaviest active neutrino masses, respectively. Since our main interest is to find a lower bound on the symmetry breaking scale for lepton number, for the rest of our calculations we set $\varepsilon_1$ to its maximal allowed value for a given $M_{N_1}$. Therefore, the only information we need from the active neutrino sector are the mass splittings and we use the central values listed in Ref.~\cite{Esteban:2018azc}
\begin{align}
\Delta m^2_{31} & = 2.528 \times 10^{-3} \,\, {\rm eV}^{2},
\end{align}
since we work with the normal hierarchy scenario for the active neutrino masses. We find that our conclusions do not change for inverted hierarchy.

The bound in Eq.~\eqref{eq:epsilonDI} implies a lower bound on the mass of the lightest right-handed neutrino of $M_{N_1} \gtrsim 10^8$ GeV \cite{Davidson:2002qv, Hambye:2003rt} in order to achieve the measured baryon asymmetry. The addition of flavor effects can lower this bound by an order of magnitude (see e.g. \cite{Davidson:2008bu,Blanchet:2012bk}); furthermore, if there are large cancellations in the tree-level and one-loop contribution to the neutrino masses the mass can be lowered to $10^6$ GeV~\cite{Moffat:2018wke}. In this work we will focus on the hierarchical spectrum.

One has the freedom to fix the value of $M_{N_1}$ and show results for different values of $\widetilde{m}_1$ in  Fig.~\ref{fig:m1tilde}. Since the decay width of $N_1$ is proportional to this parameter, the smaller it is, the more $N_1$ departs from equilibrium. Our results in Fig.~\ref{fig:m1tilde} demonstrate that for $\widetilde{m}_1<10^{-5}$ the scattering interaction goes out-of-equlibrium before the decay rate thermalizes. At this point the $N_1$ abundance plateaus and once the  decay rate enters into thermal equilibrium it starts decreasing. A smaller value for $\widetilde{m}_1$ implies that the asymmetry starts to be generated at a later time but it saturates at the same value, so the asymmetry becomes independent of this parameter (as long as $N_1$ thermalizes). For these small values of $\widetilde{m}_1$ the asymmetry depends on the temperature at which $N_1$ departs from equilibrium rather than by how much it departs from equilibrium, so it depends on the strength of the scattering mediated by the $Z_L$. The dependence of the baryon asymmetry on $\widetilde{m}_1$ without the new gauge interaction has been studied in detail in previous works, see e.g.~\cite{Buchmuller:2004nz}. Furthermore, we find that the maximal baryon asymmetry is found when $\widetilde{m}_1$  is close to the equilibration mass $\tilde{m}^*$.

\begin{figure}[t]
\centering
\includegraphics[width=0.495\linewidth]{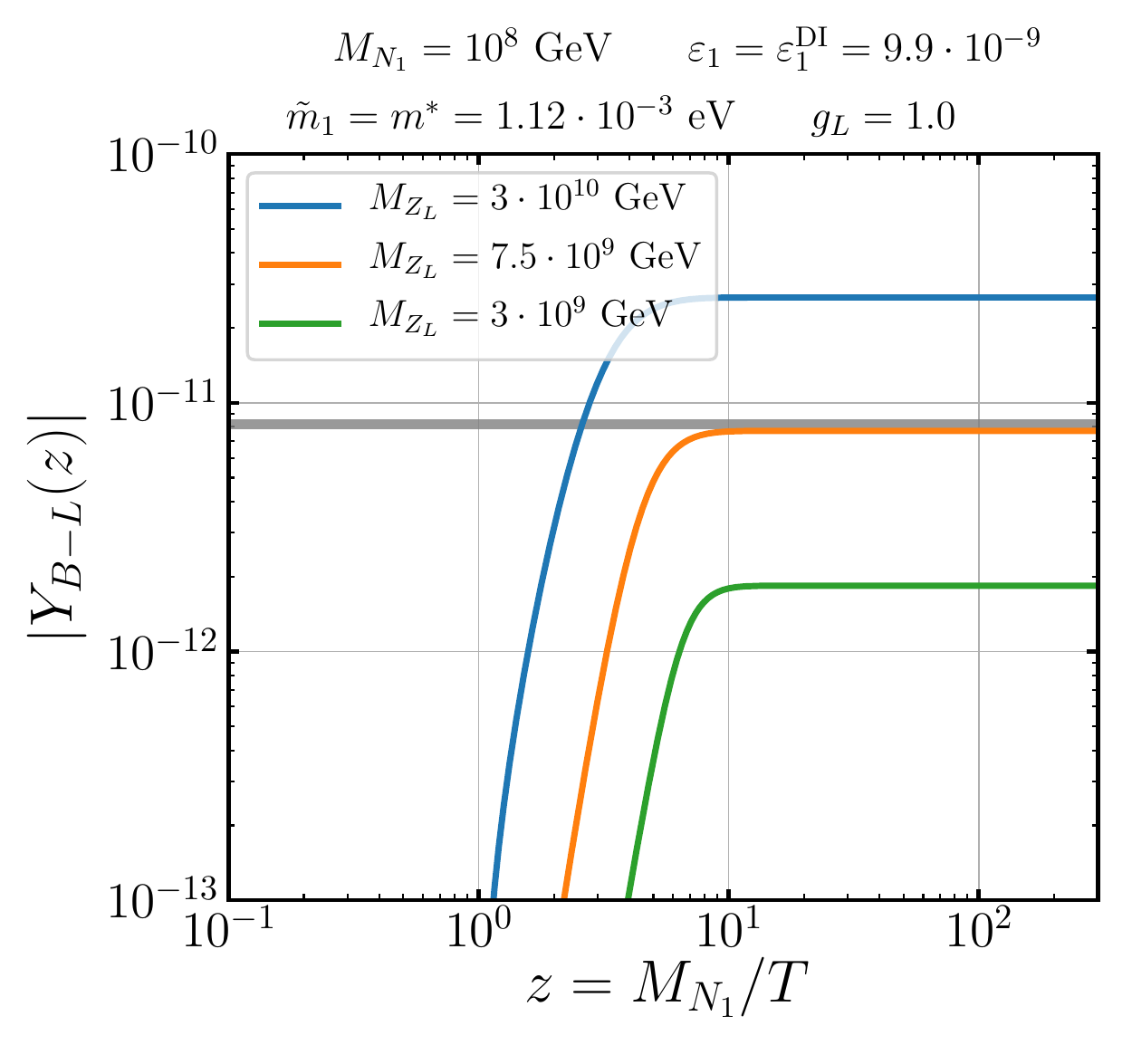}
\includegraphics[width=0.495\linewidth]{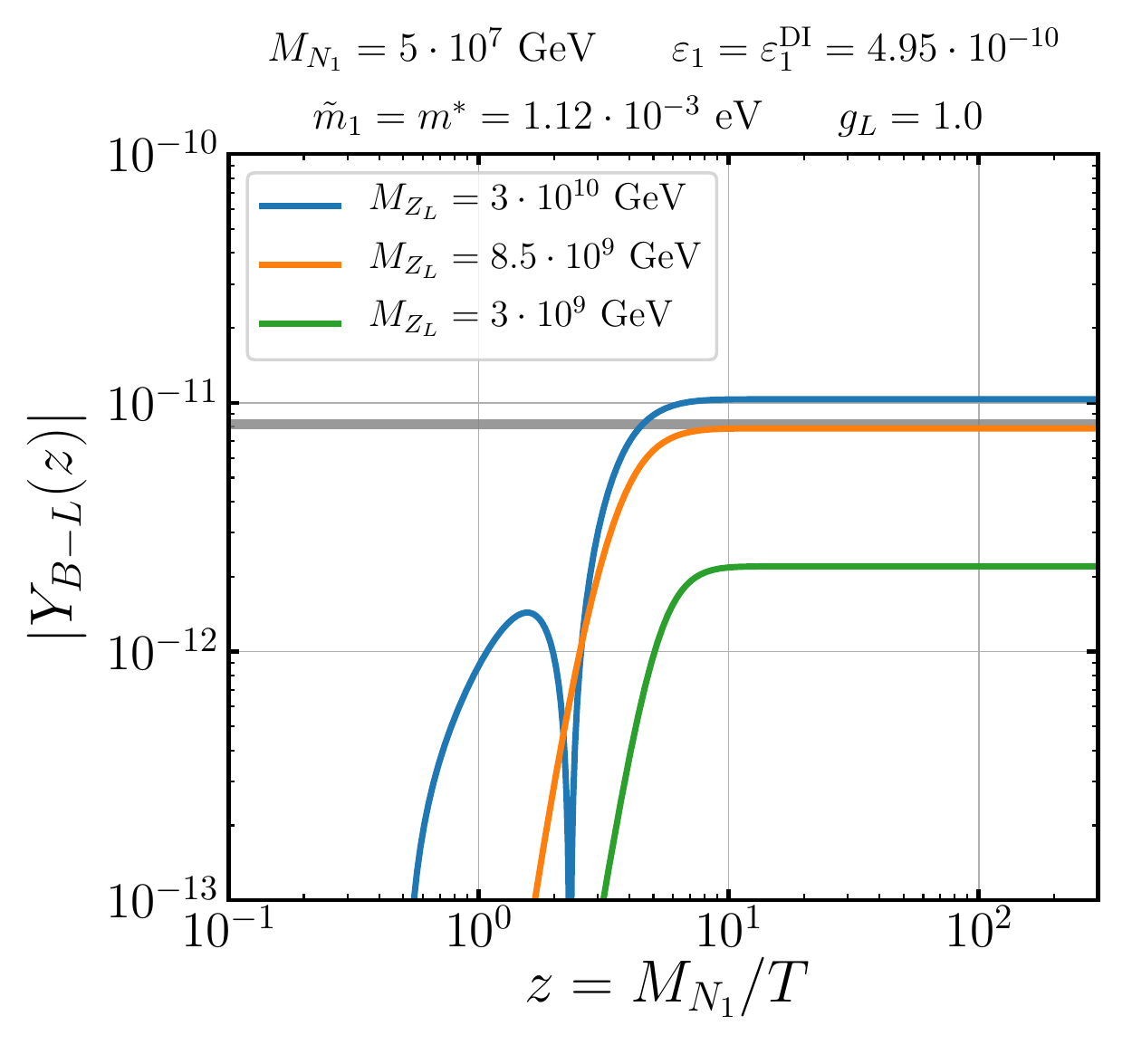}
\caption{Evolution of the $B-L$ asymmetry as a function of the parameter $z$. For the left (right) panel we have fixed $M_{N_1}=10^8$ GeV ($=5 \times 10^7$ GeV) and the other parameters as shown in the figure. Different colors correspond to different values of $M_{Z_L}$ as shown in the legend. The gray band corresponds to the measured value of the baryon asymmetry. For values of $M_{Z_L}$ smaller than $7.5\times 10^9$ GeV the mechanism produces a $B-L$ asymmetry smaller than the measured one. }
\label{upperbound}
\end{figure}
In Fig.~\ref{upperbound} we show the evolution of the $B-L$ asymmetry as a function of the parameter $z$. We fix the mass of $N_1$ and then vary the mass of $Z_L$. A lower value of $M_{Z_L}$ keeps the right-handed neutrino into thermal equilibrium for longer time so the $B-L$ asymmetry decreases. Since the cross-section scales as $(g_L/M_{Z_L})^4$, by imposing the condition to generate the observed baryon asymmetry we can find a lower bound on the ratio $M_{Z_L}/g_L$
\beq
\frac{M_{Z_L}}{g_L} \gtrsim 8 \times 10^{9} \,\, {\rm GeV} \,\,\,\, \Rightarrow \,\,\,\, v_L \gtrsim 4 \times 10^{9} \,\, {\rm GeV}.
\eeq
Notice that $M_{Z_L}=2 g_L v_L$, where $v_L$ is the vacuum expectation value of the Higgs $S_L$ breaking spontaneously $\U(1)_L$.
We find this lower bound from setting $M_{N_1}=10^8$ GeV which requires $M_{Z_L}>75 \, M_{N_1}$, and hence, our approach of integrating out $Z_L$ is well justified. However, this result applies for $g_L$ of order one because if the gauge coupling is taken to be $g_L\ll 1$ then $M_{Z_L}$ could be lower and our approach of integrating out the $Z_L$ and ignoring the Boltzmann equation for the evolution of its number density no longer works. It is possible to assume a very small $g_L$ gauge coupling but in this case it is hard to imagine a simple UV completion of the theory where the Abelian symmetries originate from non-Abelian gauge symmetries and where gauge coupling unification could be realized.

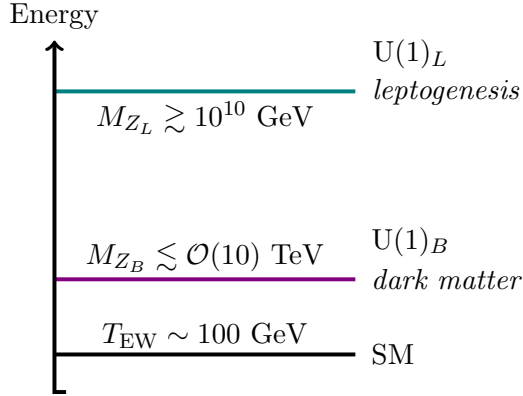
\begin{figure}[t]
\begin{center}
\begin{tikzpicture}[line width=1.5 pt,node distance=1 cm and 1.5 cm]
% \filldraw[fill=blue!20!white, draw=none] (0,0.5) rectangle (4,4);%sphalerons?
\node (A) at (0, 0) {};
\node (B) at (0, 5) {${\rm Energy}$};
\node (C) at (4.5, 0.5) {$\text{SM}$};
\node (D) at (5.2, 1.5) {$\textit{dark matter}$};
\node (E) at (5.2, 4) {$\textit{leptogenesis}$};
\node (F) at (2, 3.65) {$M_{Z_L}\gtrsim 10^{10} \text{ GeV}$};
\node (G) at (2, 1.8) {$M_{Z_B}\lesssim {\cal O}(10) \text{ TeV}$};
\node (H) at (2, 0.75) {$T_\text{EW} \sim 100 \text{ GeV}$};
\node (I) at (4.7, 2.0) {$\U(1)_B$};
\node (J) at (4.7, 4.5) {$\U(1)_L$};
\draw[black] (0,0.5) -- (4,0.5); %SM
\draw[violet] (0,1.5) -- (4,1.5); %U1B
\draw[teal] (0,4) -- (4,4);%U1L
\draw[->, to path={-| (\tikztotarget)}]
  (A) edge (B);
\end{tikzpicture}
\end{center}
\caption{Energy scales for spontaneous $B$ and $L$ breaking. The $\U(1)_B$ scale is bounded from above by the dark matter 
constraints, while the $\U(1)_L$ is bounded from below by the leptogenesis constraints.} 
\label{fig:diagramscales}
\end{figure}

In Fig.~\ref{fig:diagramscales} we show a summary of the different energy scales for spontaneous $B$ and $L$ breaking. The $\U(1)_B$ scale is bounded from above by the dark matter 
constraints, while the $\U(1)_L$ is bounded from below by the leptogenesis constraints. Notice that the bound coming from leptogenesis is conservative and can be lowered if flavor effects are taken into account or if one considers resonant leptogenesis~\cite{Pilaftsis:2003gt} or implements the mechanism of electroweak baryogenesis~\cite{Kuzmin:1985mm,Cohen:1990it} to generate the baryon asymmetry.

As a final comment we want to point out a possible observational signature of this mechanism. The spontaneous breaking of $\U(1)_L$ at very high temperatures leads to the formation of cosmic strings that can radiate gravitational waves, and hence, could be detected as a stochastic gravitational wave background in laser interferometers. As we have discussed, successful leptogenesis requires the spontaneous breaking scale to be $v_L \simeq 10^9 - 10^{15}$ GeV and for these scales the gravitational wave signal could be observed by LISA~\cite{amaroseoane2017laser}, BBO~\cite{Crowder:2005nr}, DECIGO~\cite{Kawamura_2008} or NANOGrav~\cite{Arzoumanian:2018saf}. The authors of Ref.~\cite{Fornal:2020esl} have pointed out that in theories with spontaneous breaking of $\U(1)_B$ and $\U(1)_L$ a feature arises in the gravitational wave spectrum from a combined signal of cosmic strings and a possible strong first order phase transition around the TeV scale for $\U(1)_B$.

\FloatBarrier

%%%%%%%%%%%%%%%%%%%%%%%%%%%%%%%%%%%%%%%%%%%%%%%%%%%%%%%%%%%%%%%%%%%%%%%%%%%%%%%
\section{BARYON AND DARK MATTER ASYMMETRIES}
\label{sec:models}
%%%%%%%%%%%%%%%%%%%%%%%%%%%%%%%%%%%%%%%%%%%%%%%%%%%%%%%%%%%%%%%%%%%%%%%%%%%%%%%
\begin{figure}[b]
\centering
\includegraphics[width=0.52\linewidth]{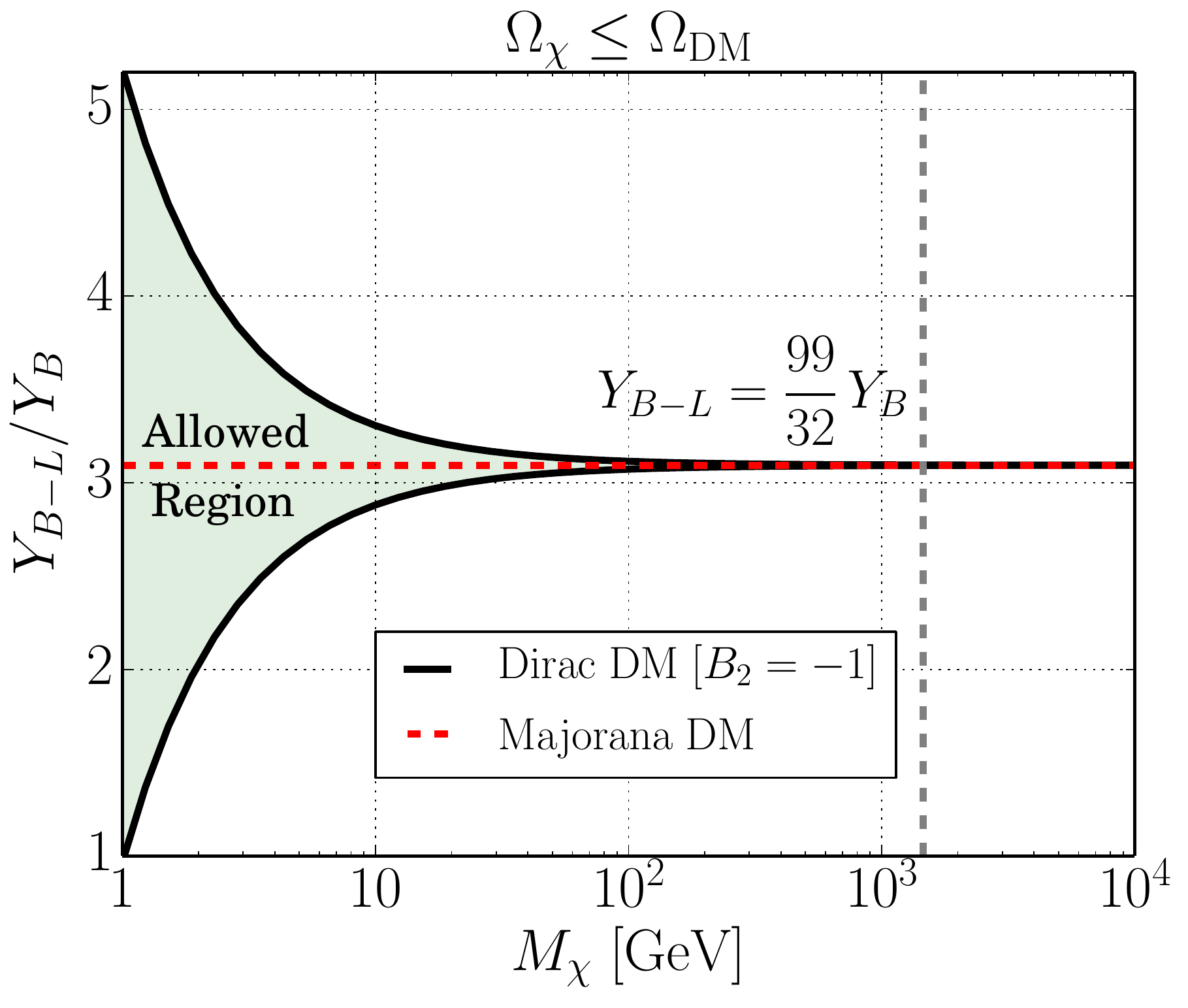}
\caption{Relation between the $B-L$ and baryon asymmetries in the two models discussed in the text. In the model with Majorana dark matter 
the relation is determined by the dashed red line, while in the model with Dirac dark matter the same relation is bounded by the black lines that correspond to $\Omega_{\rm DM} = 0.12$. The area shaded in green corresponds to $\Omega_{\rm DM}< 0.12$. The vertical dashed line corresponds to the Xenon-1T direct detection bound for a benchmark value of the gauge coupling $g_B = 0.3$ (for details, see Ref.~\cite{FileviezPerez:2018jmr}).}
\label{leptonbaryon}
\end{figure}
In the previous section we discussed how to generate the $B-L$ asymmetry through the out-of-equilibrium decays of the right-handed neutrinos in a theory with spontaneous $L$ breaking at the high scale. As we mentioned above, one can have simple theories where one can understand the spontaneous $B$ and $L$ breaking and here we will show how the $B-L$ asymmetry generated through leptogenesis is converted into a baryon asymmetry. 
We investigate two main scenarios: \textit{a)} The lepton asymmetry is transferred into a baryon asymmetry when the DM candidate is a Majorana fermion, and 
\textit{b)} The lepton and dark matter asymmetries contribute to the baryon asymmetry when the DM is a Dirac fermion.

\begin{itemize}

\item {\textit{Majorana DM: Lepton Asymmetry converted into Baryon Asymmetry}}

In the theory proposed in Ref.~\cite{Perez:2014qfa} one can define an anomaly-free theory based on $\SU(3)_C \otimes \SU(2)_L \otimes \U(1)_Y \otimes  \U(1)_B \otimes \U(1)_{L}$ 
adding only four fermionic representations:
\begin{center}
$\Psi_L {\sim \left(\mathbf{1}, \, \mathbf{2}, \,  \tfrac{1}{2}, \, \tfrac{3}{2}, \tfrac{3}{2} \right)}$, $\Psi_R \sim \left(\mathbf{1}, \, \mathbf{2}, \, \tfrac{1}{2}, \, -\tfrac{3}{2}, -\tfrac{3}{2} \right)$, \\[1.5ex]
$\Sigma_L  \sim (\mathbf{1}, \, \mathbf{3}, \, 0, \, -\tfrac{3}{2}, -\tfrac{3}{2})$ and $ \chi_L \sim (\mathbf{1}, \, \mathbf{1}, \, 0, \, -\tfrac{3}{2}, -\tfrac{3}{2})$, 
\end{center}
plus three right-handed neutrinos, $\nu_R \sim ({\bf 1},{\bf 1},0,0,1)$. The new fermions acquire mass after the spontaneous breaking of $\U(1)_B$ and Majorana neutrino masses 
through the seesaw mechanism can be generated when $\U(1)_L$ is broken in two units. The new Higgs $S_B \sim (\mathbf{1}, \mathbf{1}, 0, 3, 3)$ can generate masses 
for the extra fermions using the following Yukawa interactions~\cite{Perez:2014qfa}:
\begin{eqnarray}
- \mathcal{L}_M &\supset& y_1 \overline{\Psi}_R H \chi_L  + y_2 H^\dagger \Psi_L \chi_L + y_3 H^\dagger \Sigma_L \Psi_L + y_4 \overline{\Psi}_R \Sigma_L H  \nonumber \\
&& + y_\psi \overline{\Psi}_R  {\Psi}_L S_B^* + y_\chi \chi_L  \chi_L S_B + y_\Sigma  \textrm{Tr} \Sigma_L^2  S_B + \textrm{h.c.} \,.
\end{eqnarray}
Recently, we have investigated the phenomenological and cosmological aspects of this theory in Refs.~\cite{FileviezPerez:2019jju,FileviezPerez:2020gfb,Perez:2020baq}. In these studies we have shown that in order to satisfy the dark matter relic density constraints the symmetry breaking scale for $\U(1)_B$ must be below the multi-TeV scale.
It is important to mention that in this theory the DM candidate is automatically stable and predicted to be a Majorana fermion. In this theory the $B-L$ asymmetry can be 
converted with a similar conversion factor even assuming that $\U(1)_B$ is broken close to the electroweak scale~\cite{Perez:2014qfa}:
\begin{equation}
Y_B = \frac{32}{99} \, Y_{B-L} \approx 0.32 \, Y_{B-L}.
\end{equation} 
Notice that the conversion factor is different from the one in the Standard Model, i.e. smaller than $28/79 \approx 0.35$. Since dark matter is a Majorana fermion there is no asymmetry in the dark matter sector.

\item  {\textit{Dirac DM: Lepton and DM Asymmetries converted into Baryon Asymmetry}}

In the theory proposed in Ref.~\cite{Duerr:2013dza} one can cancel all $B$ and $L$ gauge anomalies by adding six representations:
\begin{eqnarray*}
\Psi_L & \sim& ({\bf 1}, {\bf 2}, -\tfrac{1}{2}, B_1, L_1), \quad \quad \quad  \Psi_R  \sim ({\bf 1}, {\bf 2}, -\tfrac{1}{2}, B_2, L_2),\\
\eta_L & \sim & ({\bf 1}, {\bf 1}, -1, B_2, L_2), \quad \quad \quad    \,  \, \eta_R \sim ({\bf 1}, {\bf 1}, -1, B_1, L_1),\\
\chi_L & \sim & ({\bf 1}, {\bf 1}, 0, B_2, L_2), \quad \quad \quad \quad   \chi_R \sim ({\bf 1}, {\bf 1}, 0, B_1, L_1),
\end{eqnarray*}
where $B_2 - B_1 = 3$ and $L_2 - L_1 = 3$ are fixed by anomaly cancellation. We note that in the context of this theory one has more freedom, although less predictability, to choose the baryon and lepton charges. 
In the context of this scenario, the dark matter can be either Dirac or Majorana, depending on the charge assignment, being in general the former case, while its Majorana nature 
specifically requires $B_2 (L_2) = -B_1(L_1) = 3/2$. See Refs.~\cite{Duerr:2013lka,Duerr:2014wra,FileviezPerez:2018jmr} for the study of the dark matter properties in this context.

As in the previous case, $S_B \sim ({\bf 1}, {\bf 1}, 0 , 3, 3)$ is needed to generate mass for the anomalons through the following Yukawa interactions:
\begin{eqnarray}
- \mathcal{L}_{II} &\supset& y_1 \bar \Psi_L H \eta_R + y_2 \bar \Psi_R H  \eta_L + y_3 \bar \Psi_L \tilde{H} \chi_R + y_4 \bar \Psi_R \tilde{H} \chi_L \nonumber \\
& + & y_\Psi \bar \Psi_L \Psi_R S_B^* + y_\eta \bar \eta_R \eta_L S_B^* + y_\chi \bar \chi_R \chi_L S_B^* +  \rm{h.c.}\, .
\end{eqnarray}
In Ref.~\cite{Perez:2013tea} the authors studied the relation between the $B-L$, dark matter and baryon asymmetries 
when the $\U(1)_B$ symmetry is broken at a scale close to the electroweak scale. The `t Hooft operator associated to the sphaleron processes in this theory has to respect local baryon number and it is given by,
\beq
\left( Q_L Q_L Q_L \ell_L \right)^3 \bar \Psi_R \Psi_L,
\eeq
using this and the relations between the different chemical potentials, the authors showed that the baryon asymmetry is related to the$B-L$ and dark matter asymmetries as follows 
\begin{equation}
Y_B = \frac{32}{99} Y_ {B-L} + \frac{(15-14B_2)}{198} Y_{\rm DM},
\end{equation}
where $B_2$ is fixed by the particular charge assignment. Here $Y_{\rm DM}$ is the asymmetry in the dark matter sector.
Notice that in this context the $B-L$ and dark matter asymmetries are related to the baryon asymmetry by sphaleron processes. 

In Fig.~\ref{leptonbaryon} we show the relation between the $B-L$ and baryon asymmetries in the models discussed above. The figure displays the necessary amount of $B-L$ asymmetry in order to generate the observed matter-antimatter asymmetry in the theories presented.
In the model with Majorana dark matter the relation is determined by the dashed red line.  In the model with Dirac dark matter, we impose the dark matter asymmetry to satisfy the constraint constraint $\Omega_\chi \leq \Omega_\text{DM}$, where $\Omega_{\rm DM} h^2=0.12$~\cite{Aghanim:2018eyx}. The bound on the $B-L$ asymmetry is shown by the black lines, and we assume $B_2=-1$ for illustration. Notice that in the model with Dirac dark 
matter the relation changes as a function of the dark matter mass. In this case one also needs to make sure that the symmetric component of the dark matter relic density is small making sure that one has a large annihilation cross section such as $\chi \bar{\chi} \to Z_B h_B$~\cite{FileviezPerez:2018jmr}. Now, in order to satisfy the bound coming from direct detection 
we need to assume that the dark matter mass is larger than approximately 1 TeV, for details see Ref.~\cite{FileviezPerez:2018jmr}.
It is important to mention that the mechanism for asymmetric dark matter present in the model with a Dirac dark matter candidate 
is quite unique, since the sphaleron processes convert partially the $B-L$ and dark matter asymmetries into a baryon asymmetry.
\end{itemize}

%%%%%%%%%%%%%%%%%%%%%%%%%%%%%%%%%%%%%%%%%%%%%%%%%%%%%%%%%%%%%%%%%%%%%%%%%%%%%%%
\section{DARK MATTER ASYMMETRY}
\label{sec:DM}
%%%%%%%%%%%%%%%%%%%%%%%%%%%%%%%%%%%%%%%%%%%%%%%%%%%%%%%%%%%%%%%%%%%%%%%%%%%%%%%
As we demonstrated in the previous section, in theories with gauge baryon and lepton numbers, the baryon asymmetry is related to the $B-L$ and the dark matter asymmetry.
Such primordial asymmetries are respected by all processes (perturbative and non-perturbative) from the scale where they are generated until today, and any of them, or both, will generate a baryon asymmetry. In Sec.~\ref{sec:leptogenesis}, we discussed leptogenesis as the main mechanism in the context of gauged baryon and lepton numbers to generate a $B-L$ asymmetry. We remark that the $\U(1)_B$ gauge symmetry imposes the initial baryon asymmetry to be zero.  

\begin{figure}[b]
\centering
\begin{eqnarray*}
&& \quad \quad  \begin{gathered}
\resizebox{.25\textwidth}{!}{%
\begin{tikzpicture}[line width=1.5 pt,node distance=1 cm and 1.5 cm]
\coordinate[label = left: $N_1$] (p1);
\coordinate[right = of p1](p2);
\coordinate[above right = of p2, label=right: $ \phi$] (v1);
\coordinate[below right = of p2,label=right: $ \chi_L$] (v2);
\draw[fermionnoarrow] (p1) -- (p2);
\draw[scalar] (v1) -- (p2);
\draw[fermion] (p2) -- (v2);
\draw[fill=black] (p2) circle (.07cm);
\end{tikzpicture}}
\end{gathered}
 \quad \quad +   \quad
\begin{gathered}
\resizebox{.32\textwidth}{!}{%
\begin{tikzpicture}[line width=1.5 pt,node distance=1 cm and 1.5 cm]
\coordinate[label = left: $N_1$] (p1);
\coordinate[right = 0.8cm  of p1](p1a);
\coordinate[right = of p1a](p1b);
\coordinate[right = 0.9 cm of p1b](p2);
\coordinate[above right = of p2, label=right: $ \phi$] (v1);
\coordinate[below right = of p2,label=right: $ \chi_L$] (v2);
\coordinate[above right = 1cm of p1a,label=$\ell_L$](vaux1);
\coordinate[below right = 1cm of p1a,label=$H$](vaux2);
\coordinate[right=0.5cm of p1b,label=above:$N_{j\neq1}$](vaux3);
\draw[fermionnoarrow] (p1) -- (p1a);
\draw[scalar] (v1) -- (p2);
\draw[fermion] (p2) -- (v2);
\draw[fermionnoarrow] (p1b)--(p2);
\semiloop[fermionnoarrow]{p1a}{p1b}{0};
\hello[scalarnoarrow]{p1a}{p1b}{0};
\draw[fill=black] (p2) circle (.07cm);
\draw[fill=black] (p1a) circle (.07cm);
\draw[fill=black] (p1b) circle (.07cm);
\end{tikzpicture}}
\end{gathered} % \!\!\! + \quad 
\\
&&+ \quad 
\begin{gathered}
\resizebox{.32\textwidth}{!}{%
\begin{tikzpicture}[line width=1.5 pt,node distance=1 cm and 1.5 cm]
\coordinate[label = left: $N_1$] (p1);
\coordinate[right = 0.8cm  of p1](p1a);
\coordinate[right = of p1a](p1b);
\coordinate[right = 0.9 cm of p1b](p2);
\coordinate[above right = of p2, label=right: $ \phi$] (v1);
\coordinate[below right = of p2,label=right: $ \chi_L$] (v2);
\coordinate[above right = 1cm of p1a,label=$\chi_L$](vaux1);
\coordinate[below right = 1cm of p1a,label=$\phi$](vaux2);
\coordinate[right=0.5cm of p1b,label=above:$N_{j\neq1}$](vaux3);
\draw[fermionnoarrow] (p1) -- (p1a);
\draw[scalar] (v1) -- (p2);
\draw[fermion] (p2) -- (v2);
\draw[fermionnoarrow] (p1b)--(p2);
\semiloop[fermionnoarrow]{p1a}{p1b}{0};
\hello[scalarnoarrow]{p1a}{p1b}{0};
\draw[fill=black] (p2) circle (.07cm);
\draw[fill=black] (p1a) circle (.07cm);
\draw[fill=black] (p1b) circle (.07cm);
\end{tikzpicture}}
\end{gathered}  \!\!\! + \quad 
\begin{gathered}
\resizebox{.3\textwidth}{!}{%
\begin{tikzpicture}[line width=1.5 pt,node distance=1 cm and 1.5 cm]
\coordinate[label = left: $N_1$] (p1);
\coordinate[right = of p1](p2);
\coordinate[right=0.55 cm of p2](vmare);
\coordinate[above=0.5cm of vmare,label=$\chi_L$](vaux1);
\coordinate[below=1.1cm of vmare,label=$\phi$](vaux2);
\coordinate[right= 1cm of vmare,label=right:$N_{j\neq1}$](vaux3);
\coordinate[above right = of p2](v1a);
\coordinate[right = of v1a, label=right: $\phi$] (v1);
\coordinate[below right = of p2](v2a);
\coordinate[right = of v2a,label=right: $ \chi_L$] (v2);
\draw[fermionnoarrow] (p1) -- (p2);
\draw[scalar] (v1) -- (v1a);
\draw[fermionnoarrow] (v1a)--(v2a);
\draw[fermion](v1a)--(p2);
\draw[scalar] (p2)--(v2a);
\draw[fermion] (v2a) -- (v2);
\draw[fill=black] (p2) circle (.07cm);
\draw[fill=black] (v1a) circle (.07cm);
\draw[fill=black] (v2a) circle (.07cm);
\end{tikzpicture}}
\end{gathered}
\end{eqnarray*}
\caption{Feynman diagrams for the decays of the lightest right-handed neutrino $N_1$ that contribute to $\varepsilon_{\rm DM}$.}
\label{fig:DMdecay}
\end{figure}
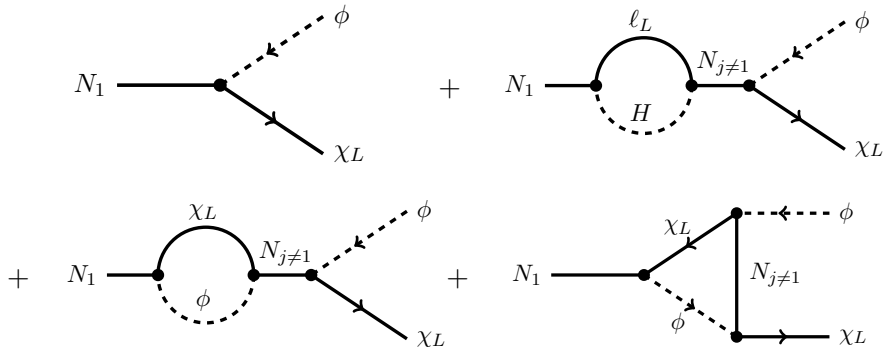

In this section, we implement the mechanism proposed in Ref.~\cite{Falkowski:2011xh} to generate a dark matter asymmetry in the scenario with Dirac dark matter.  We make use of the same ingredients responsible for leptogenesis: CP violation from the neutrino sector and the out-of-equilibrium of the right-handed neutrinos. The only new degree of freedom that needs to be introduced is a complex scalar $\phi$ with quantum numbers 
\beq
\phi \sim (\mathbf{1}, \, \mathbf{1}, \, 0, \, B_2, \, L_2-1),
\eeq
which does not take a vev, and hence, does not mix with the other scalars in the theory. The following Yukawa interactions can be added to the Lagrangian density
\beq
-\mathcal{L} \supset y^i_{\rm DM}  \, \bar \chi_L \, \nu_{R}^i \,  \phi + y_{L}^i \overline{\ell}_{L}^i \Psi_R \phi^*  + y_{R}^i \, \overline{e}_{R}^i \, \eta_L \phi^* + {\rm h.c.} \,,
\label{eq:YukawaDM}
\eeq
where the new Yukawa couplings are vectors in flavor space.
As a consequence of its quantum numbers, the new scalar $\phi$ and all the anomaly-canceling fermions, including the DM candidate $\chi_L$, enjoy a $Z_2$ symmetry, which ensures the stability of DM as long as $M_\phi > M_\chi$.
The scalar $\phi$ has the two-body decays $\phi \to \overline{\ell}_L \Psi_R$ and $\phi \to \overline{e}_R \, \eta_L$, and a three-body decay $\phi \to \overline{\ell}_L \, H^\dagger \, \chi_L$, mediated by a virtual $N_i$.

The CP asymmetry in the dark matter sector is defined by
\beq
\varepsilon_{\rm DM} = \frac{\Gamma(N_1 \to \phi \, \overline{\chi}_L) - \Gamma(N_1 \to \phi^* \chi_L) }{\Gamma(N_1 \to \phi \, \overline{\chi}_L) + \Gamma(N_1 \to \phi^* \chi_L)},
\eeq
and it is generated by the interference between the tree level and one-loop diagrams in Fig.~\ref{fig:DMdecay}. 

Finally, the out-of-equilibrium decays of the right-handed neutrinos through Eq.~\eqref{eq:YukawaDM} violate the global symmetry $\U(1)_\chi$ associated with the dark matter asymmetry and, together with the CP violation stated above, they will lead to a non-zero dark matter asymmetry. The Boltzmann equations that give the evolution of the $B-L$ and DM asymmetries are given by
\begin{align}
\frac{ dY_{N_1}} {dz} \simeq &  - \frac{z}{ s H(M_{N_1})} \left[ \left( \frac{Y_{N_1}}{Y^{\rm eq}_{N_1}} - 1 \right)  \gamma_{D}    + \left( \left( \frac{Y_{N_1}}{Y^{\rm eq}_{N_1}} \right)^2 -1 \right) \left(  \gamma_1 + \gamma_2  \right)  \right], \\[2ex]
\frac{ dY_{B-L}} {dz} \simeq & - \frac{z}{ s H(M_{N_1})} \left[ \frac{1}{2} \frac{Y_{B-L}}{Y_\ell^{\rm eq}} - \varepsilon_1  \left(\frac{Y_{N_1}}{Y^{\rm eq}_{N_1}} - 1 \right) \right] 2 \, \text{Br} ( N_1 \to \ell_L H ) \, \gamma_D, \\[2ex]
\frac{ dY_{\rm DM}} {dz} \simeq & - \frac{z}{ s H(M_{N_1})}  \left[ \frac{1}{2} \frac{Y_{\rm DM}}{Y_\chi^{\rm eq}} - \varepsilon_{\rm DM}  \left(\frac{Y_{N_1}}{Y^{\rm eq}_{N_1}} - 1\right) \right] 2 \, \text{Br} ( N_1 \to \chi_L \phi^* ) \, \gamma_D , 
\end{align}
where we assume that $M_{\rm DM} \ll M_\phi \lesssim M_{N_1}$ and we neglect scattering processes that could change $B-L$ or DM. 
Note that in general, since one would expect $\phi$ to have a large mass or the new Yukawa couplings to be small, then $\text{Br}( N_1 \to \chi_L \, \phi^*) \lesssim \text{Br}(N_1 \to \ell_L H)$, so that the results for the $B-L$ asymmetry will not change drastically with respect to the analysis done in Sec.~\ref{sec:leptogenesis}. We note that, as discussed in Sec.~\ref{sec:models}, this dark matter asymmetry will contribute to the baryon asymmetry through the sphalerons respecting the $\U(1)_B$ symmetry.

This mechanism also produces an asymmetry in the $\phi$ field. However, this asymmetry can be washed out by scattering processes that have $\Delta \phi = -1$ and $\Delta \phi = -2$, such as $\phi \, Z_B \to \overline{\ell}_L \Psi_R$ and $\phi \, \overline{\ell}_L\to \phi^* \ell_L$, if the new Yukawa couplings $y_{L/R}^i$ are large enough to bring these processes into thermal equilibrium. We leave a detailed study of this mechanism for future work.

\clearpage
%%%%%%%%%%%%%%%%%%%%%%%%%%%%%%%%%%%%%%%%%%%%%%%%%%%%%%%%%%%%%%%%%%%%%%%%%%%%%%%
\section{SUMMARY}
\label{sec:summary}
%%%%%%%%%%%%%%%%%%%%%%%%%%%%%%%%%%%%%%%%%%%%%%%%%%%%%%%%%%%%%%%%%%%%%%%%%%%%%%%
The origin of the baryon asymmetry in the Universe is an open problem for which we must understand how baryon and lepton number are broken in nature.
We have discussed simple theories where it is possible to understand the spontaneous breaking 
of $B$ and $L$ numbers. In this context, $B$ and $L$ are local gauge symmetries and predict a dark matter candidate from the cancellation of gauge anomalies.
The breaking scale for $\U(1)_B$ is bounded from above by the dark matter relic density constraints.
In this context, the properties of the predicted dark matter candidate are crucial to understand how the baryogenesis mechanism can work.

We have investigated the mechanism of leptogenesis in theories for $B$ and $L$ spontaneous breaking. Since at very high temperatures $\U(1)_B$ 
and $\U(1)_L$ are conserved it is not possible to have any primordial lepton or baryon asymmetries. Once $\U(1)_L$ is broken, the right-handed 
neutrinos acquire masses and can generate a lepton asymmetry from their out-of-equilibrium decays. In this scenario the right-handed 
neutrinos have an additional gauge interaction with the gauge boson, $Z_L$, associated to $\U(1)_L$. When the interaction between 
the right-handed neutrinos and the $Z_L$ is large, then the right-handed neutrinos come into thermal equilibrium 
very early. We numerically solved the Boltzmann equations and discussed the correlation between the scattering processes mediated by the gauge interactions and the decays. Our numerical results show that there is a conservative 
lower bound on the $\U(1)_L$ breaking scale, i.e. $M_{Z_L}/g_L \gtrsim 8 \times 10^{9} \,\, {\rm GeV} $.

As we discussed previously, one can consider two simple theories for spontaneous $B$ and $L$ breaking. 
In one of these theories, the dark matter candidate is a Majorana fermion and the baryon
asymmetry is generated after the lepton asymmetry is converted by the sphalerons, regardless of whether $\U(1)_B$ is broken above or below the electroweak scale. In the second class of these theories, the dark matter can be a Dirac fermion 
and we have discussed a mechanism in which the out-of-equilibrium decays of the right-handed neutrinos produce $B\!-\!L$ and  dark matter asymmetries. These asymmetries are then converted into a baryon asymmetry through sphaleron processes.
These results show that theories for spontaneous $B$ and $L$ breaking provide a solution to the baryon asymmetry, the dark matter abundance and the origin of neutrino masses.

\vspace{0.5cm}
{\textit{Acknowledgments}}: The work of P.F.P. has been supported by the U.S. Department of Energy, Office of Science, Office of
High Energy Physics, under Award Number DE-SC0020443.
The work of C.M. is supported by the U.S. Department of Energy, Office of Science, Office of High Energy Physics, under Award Number DE-SC0011632 and by the Walter Burke Institute for Theoretical Physics. 

\pagebreak 

\appendix
%%%%%%%%%%%%%%%%%%%%%%%%%%%%%%%%%%%%%%%%%%%%%%%%%%%%%%%%%%%%%%%%%%%%%%%%%%%%%%%
\section{FEYNMAN GRAPHS}
%%%%%%%%%%%%%%%%%%%%%%%%%%%%%%%%%%%%%%%%%%%%%%%%%%%%%%%%
In this Appendix we list the Feynman diagrams for all processes that enter into the Boltzmann equations of the lightest right-handed neutrino, $N_1$, and the $B-L$ asymmetry, described in Eqs.~\eqref{eq:Boltzmann1}-\eqref{eq:Boltzmann2} from Sec.~\ref{sec:leptogenesis}. Those enclosed in a blue rectangle give the dominant contributions and are the ones we consider in our numerical calculations by solving Eqs.~\eqref{eq:Boltzmann3}-\eqref{eq:Boltzmann4}. We represent by a blue dot a gauge vertex interaction and by a red dot those corresponding to the Dirac neutrino Yukawa coupling.
%%%%%%%%%%%%%%%%%%%%%%%
%
\begin{eqnarray*}
%%%%%%%%%%%%%%%%%%%%%%%%%%%%%%%%%%
%
\gamma_D &:& \quad 
\begin{gathered}
\begin{tikzpicture}[line width=1.5 pt,node distance=1 cm and 1.5 cm]
\coordinate[label = left: $N$] (p1);
\coordinate[right = of p1](p2);
\coordinate[above right = of p2, label=right: $ H$] (v1);
\coordinate[below right = of p2,label=right: $ \ell_L$] (v2);
\draw[fermionnoarrow] (p1) -- (p2);
\draw[scalar] (p2) -- (v1);
\draw[fermion] (p2) -- (v2);
\draw[fill=red] (p2) circle (.1cm);
\end{tikzpicture}
\end{gathered}\\
\gamma_1&:&\quad
\begin{gathered}
\begin{tikzpicture}[line width=1.5 pt,node distance=1 cm and 1.5 cm]
\coordinate[label = left: $N$] (i1);
\coordinate[below right = 1cm of i1](v1);
\coordinate[below left = 1cm of v1, label= left:$N$](i2);
\coordinate[above right = 1cm of v1, label=right: $\ell_L$] (f1);
\coordinate[below right =  1cm of v1,label=right: $ \ell_L$] (f2);
\draw[fermionnoarrow] (i1) -- (v1);
\draw[fermionnoarrow] (i2) -- (v1);
\draw[fermion] (v1) -- (f1);
\draw[fermion] (f2) -- (v1);
\draw[fill=gray] (v1) circle (.3cm);
\end{tikzpicture}
\end{gathered} 
%%%%%%%%%%
=
%%%%%%%%%%{double,ultra thick,draw=red, top color=blue, rounded corners}
\begin{gathered}
\begin{tikzpicture}[line width=1.5 pt,node distance=1 cm and 1.5 cm,framed,background rectangle/.style={draw=blue, thick,rounded corners}]
\coordinate[label =left: $N$] (i1);
\coordinate[below right= 1cm of i1](v1);
\coordinate[ right= 0.5cm of v1,label= below:$Z_L$](vaux);
\coordinate[below left= 1cm of v1, label= left:$N$](i2);
\coordinate[right = 1 cm of v1](v2);
\coordinate[above right = 1 cm of v2, label=right: $\ell_L$] (f1);
\coordinate[below right =  1 cm of v2,label=right: $\ell_L$] (f2);
\draw[fermionnoarrow] (i1) -- (v1);
\draw[fermionnoarrow] (i2) -- (v1);
\draw[vector] (v1) -- (v2);
\draw[fermion] (v2) -- (f1);
\draw[fermion] (f2) -- (v2);
\draw[fill=cyan] (v1) circle (.1cm);
\draw[fill=cyan] (v2) circle (.1cm);
\end{tikzpicture}
\end{gathered} 
+
\begin{gathered}
\begin{tikzpicture}[line width=1.5 pt,node distance=1 cm and 1.5 cm]
\coordinate[label =left: $N$] (i1);
\coordinate[right= 1cm of i1](v1);
\coordinate[below= 0.5cm of v1,label= left:$H$](vaux);
\coordinate[right = 1cm of v1, label= right:$\ell_L$](f1);
\coordinate[below = 1 cm of v1](v2);
\coordinate[left = 1 cm of v2, label=left: $N$] (i2);
\coordinate[right =  1 cm of v2,label=right: $ \ell_L$] (f2);
\draw[fermionnoarrow] (i1) -- (v1);
\draw[fermionnoarrow] (i2) -- (v2);
\draw[scalar] (v1) -- (v2);
\draw[fermion] (v1) -- (f1);
\draw[fermion] (f2) -- (v2);
\draw[fill=red] (v1) circle (.1cm);
\draw[fill=red] (v2) circle (.1cm);
\end{tikzpicture}
\end{gathered} 
+
\begin{gathered}
\begin{tikzpicture}[line width=1.5 pt,node distance=1 cm and 1.5 cm]
\coordinate[label =left: $N$] (i1);
\coordinate[right= 1cm of i1](v1);
\coordinate[below= 0.5cm of v1,label= left:$ H$](vaux);
\coordinate[right = 1cm of v1, label= right:$\ell_L$](f1);
\coordinate[below = 1 cm of v1](v2);
\coordinate[left = 1 cm of v2, label=left: $N$] (i2);
\coordinate[right =  1 cm of v2,label=right: $ \ell_L$] (f2);
\draw[fermionnoarrow] (i1) -- (v1);
\draw[fermionnoarrow] (i2) -- (v2);
\draw[scalar] (v2) -- (v1);
\draw[fermionuchannel1] (v2) -- (f1);
\draw[fermionuchannel2] (f2) -- (v1);
\draw[fill=red] (v1) circle (.1cm);
\draw[fill=red] (v2) circle (.1cm);
\end{tikzpicture}
\end{gathered} \\
%%%%%%%%%%%%%%%%%%%%%%%%%%%%%%%%%%
\gamma_2&:&\quad
\begin{gathered}
\begin{tikzpicture}[line width=1.5 pt,node distance=1 cm and 1.5 cm,framed,background rectangle/.style={draw=blue, thick,rounded corners}]
\coordinate[label =left: $N$] (i1);
\coordinate[below right= 1cm of i1](v1);
\coordinate[ right= 0.5cm of v1,label= below:$Z_L$](vaux);
\coordinate[below left= 1cm of v1, label= left:$N$](i2);
\coordinate[right = 1 cm of v1](v2);
\coordinate[above right = 1 cm of v2, label=right: $F\text{, }e_R$] (f1);
\coordinate[below right =  1 cm of v2,label=right: $ F\text{, }e_R$] (f2);
\draw[fermionnoarrow] (i1) -- (v1);
\draw[fermionnoarrow] (i2) -- (v1);
\draw[vector] (v1) -- (v2);
\draw[fermion] (v2) -- (f1);
\draw[fermion] (f2) -- (v2);
\draw[fill=cyan] (v1) circle (.1cm);
\draw[fill=cyan] (v2) circle (.1cm);
\end{tikzpicture}
\end{gathered} \\
%%%%%%%%%%%%%%%%%%%%%%%%%%%%%
\gamma_3&:&\quad
\begin{gathered}
\begin{tikzpicture}[line width=1.5 pt,node distance=1 cm and 1.5 cm]
\coordinate[label =left: $N$] (i1);
\coordinate[below right= 1cm of i1](v1);
\coordinate[ right= 0.5cm of v1,label= below:$S_L$](vaux);
\coordinate[below left= 1cm of v1, label= left:$N$](i2);
\coordinate[right = 1 cm of v1](v2);
\coordinate[above right = 1 cm of v2, label=right: $S_B$] (f1);
\coordinate[below right =  1 cm of v2,label=right: $S_B$] (f2);
\draw[fermionnoarrow] (i1) -- (v1);
\draw[fermionnoarrow] (i2) -- (v1);
\draw[scalar] (v1) -- (v2);
\draw[scalar] (v2) -- (f1);
\draw[scalar] (v2) -- (f2);
\draw[fill=black] (v1) circle (.07cm);
\draw[fill=black] (v2) circle (.07cm);
\end{tikzpicture}
\end{gathered} \\
%%%%%%%%%%%%%%%%%%%%%%%%%%%%%%%%%%
\gamma_4&:& \quad 
\begin{gathered}
\begin{tikzpicture}[line width=1.5 pt,node distance=1 cm and 1.5 cm]
\coordinate[label =left: $N$] (i1);
\coordinate[below right= 1cm of i1](v1);
\coordinate[ right= 0.5cm of v1,label= below:$ H$](vaux);
\coordinate[below left= 1cm of v1, label= left:$ \ell_L$](i2);
\coordinate[right = 1 cm of v1](v2);
\coordinate[above right = 1 cm of v2, label=right: $ t_R$] (f1);
\coordinate[below right =  1 cm of v2,label=right: $Q_L$] (f2);
\draw[fermionnoarrow] (i1) -- (v1);
\draw[fermion] (i2) -- (v1);
\draw[scalar] (v2) -- (v1);
\draw[fermion] (f1) -- (v2);
\draw[fermion] (v2) -- (f2);
\draw[fill=red] (v1) circle (.1 cm);
\draw[fill=black] (v2) circle (.07cm);
\end{tikzpicture}
\end{gathered} 
\quad =
\begin{gathered}
\begin{tikzpicture}[line width=1.5 pt,node distance=1 cm and 1.5 cm]
\coordinate[label =left: $N$] (i1);
\coordinate[below right= 1cm of i1](v1);
\coordinate[ right= 0.5cm of v1,label= below:$ H$](vaux);
\coordinate[below left= 1cm of v1, label= left:$  \ell_L$](i2);
\coordinate[right = 1 cm of v1](v2);
\coordinate[above right = 1 cm of v2, label=right: $t_R$] (f1);
\coordinate[below right =  1 cm of v2,label=right: $ Q_L$] (f2);
\draw[fermionnoarrow] (i1) -- (v1);
\draw[fermion] (v1) -- (i2);
\draw[scalar] (v1) -- (v2);
\draw[fermion] (v2) -- (f1);
\draw[fermion] (f2) -- (v2);
\draw[fill=red] (v1) circle (.1 cm);
\draw[fill=black] (v2) circle (.07cm);
\end{tikzpicture}
\end{gathered}  \\
%
%%%%%%%%%%%%%%%%%%%%%%%%%%%%%%%%%%
\gamma_5&:& \quad 
\begin{gathered}
\begin{tikzpicture}[line width=1.5 pt,node distance=1 cm and 1.5 cm]
\coordinate[label =left: $N$] (i1);
\coordinate[below right= 1cm of i1](v1);
\coordinate[above right = 1cm of v1, label= right:$ H$](f1);
\coordinate[below left = 1 cm of v1, label=left: $N$] (i2);
\coordinate[below right =  1 cm of v1,label=right: $ H$] (f2);
\draw[fermionnoarrow] (i1) -- (v1);
\draw[fermionnoarrow] (i2) -- (v1);
\draw[scalar] (f2) -- (v1);
\draw[scalar] (v1) -- (f1);
\draw[fill=gray] (v1) circle (.2cm);
\end{tikzpicture}
\end{gathered}
=
\begin{gathered}
\begin{tikzpicture}[line width=1.5 pt,node distance=1 cm and 1.5 cm]
\coordinate[label =left: $N$] (i1);
\coordinate[right= 1cm of i1](v1);
\coordinate[below= 0.5cm of v1,label= left:$\ell_L$](vaux);
\coordinate[right = 1cm of v1, label= right:$ H$](f1);
\coordinate[below = 1 cm of v1](v2);
\coordinate[left = 1 cm of v2, label=left: $N$] (i2);
\coordinate[right =  1 cm of v2,label=right: $ H$] (f2);
\draw[fermionnoarrow] (i1) -- (v1);
\draw[fermion] (v1) -- (v2);
\draw[fermionnoarrow] (i2) -- (v2);
\draw[scalar] (f2) -- (v2);
\draw[scalar] (v1) -- (f1);
\draw[fill=red] (v1) circle (.1cm);
\draw[fill=red] (v2) circle (.1cm);
\end{tikzpicture}
\end{gathered}
+
\begin{gathered}
\begin{tikzpicture}[line width=1.5 pt,node distance=1 cm and 1.5 cm]
\coordinate[label =left: $N$] (i1);
\coordinate[right= 1cm of i1](v1);
\coordinate[below= 0.5cm of v1,label= left:$\ell_L$](vaux);
\coordinate[right = 1cm of v1, label= right:$ H$](f1);
\coordinate[below = 1 cm of v1](v2);
\coordinate[left = 1 cm of v2, label=left: $N$] (i2);
\coordinate[right =  1 cm of v2,label=right: $ H$] (f2);
\draw[fermionnoarrow] (i1) -- (v1);
\draw[fermion] (v2) -- (v1);
\draw[fermionnoarrow] (v2) -- (i2);
\draw[scalaruchannel1] (f2) -- (v1);
\draw[scalaruchannel2] (v2) -- (f1);
\draw[fill=red] (v1) circle (.1cm);
\draw[fill=red] (v2) circle (.1cm);
\end{tikzpicture}
\end{gathered} 
+
\begin{gathered}
\begin{tikzpicture}[line width=1.5 pt,node distance=1 cm and 1.5 cm]
\coordinate[label =left: $N$] (i1);
\coordinate[below right= 1cm of i1](v1);
\coordinate[right= 0.5cm of v1,label= below:$S_L$](vaux);
\coordinate[right = 1 cm of v1](v2);
\coordinate[above right = 1cm of v2, label= right:$H$](f1);
\coordinate[below left = 1 cm of v1, label=left: $N$] (i2);
\coordinate[below right =  1 cm of v2,label=right: $ H$] (f2);
\draw[fermionnoarrow] (i1) -- (v1);
\draw[scalar] (v1) -- (v2);
\draw[fermionnoarrow] (v1) -- (i2);
\draw[scalar] (f2) -- (v2);
\draw[scalar] (v2) -- (f1);
\draw[fill=red] (v1) circle (.1cm);
\draw[fill=black] (v2) circle (.07cm);
\end{tikzpicture}
\end{gathered} 
 \\
%%%%%%%%%%%%%%%%%%%%%%%%%%%%%%%%%%
\gamma_{6}&:& \quad 
\begin{gathered}
\begin{tikzpicture}[line width=1.5 pt,node distance=1 cm and 1.5 cm]
\coordinate[label =left: $N$] (i1);
\coordinate[right= 1cm of i1](v1);
\coordinate[below= 0.5cm of v1,label= left:$ H$](vaux);
\coordinate[right = 1cm of v1, label= right:$\ell_L$](f1);
\coordinate[below = 1 cm of v1](v2);
\coordinate[left = 1 cm of v2, label=left: $ t_R$] (i2);
\coordinate[right = 1 cm of v2,label=right: $ Q_L$] (f2);
\draw[fermionnoarrow] (i1) -- (v1);
\draw[scalar] (v1) -- (v2);
\draw[fermion] (v2) -- (i2);
\draw[fermion] (f2) -- (v2);
\draw[fermion] (v1) -- (f1);
\draw[fill=red] (v1) circle (.1cm);
\draw[fill=black] (v2) circle (.07cm);
\end{tikzpicture}
\end{gathered}
\ = \ 
\begin{gathered}
\begin{tikzpicture}[line width=1.5 pt,node distance=1 cm and 1.5 cm]
\coordinate[label =left: $N$] (i1);
\coordinate[right= 1cm of i1](v1);
\coordinate[below= 0.5cm of v1,label= left:$ H$](vaux);
\coordinate[right = 1cm of v1, label= right:$\ell_L$](f1);
\coordinate[below = 1 cm of v1](v2);
\coordinate[left = 1 cm of v2, label=left: $Q_L$] (i2);
\coordinate[right =  1 cm of v2,label=right: $t_R$] (f2);
\draw[fermionnoarrow] (i1) -- (v1);
\draw[scalar] (v1) -- (v2);
\draw[fermion] (i2) -- (v2);
\draw[fermion] (v2) -- (f2);
\draw[fermion] (v1) -- (f1);
\draw[fill=red] (v1) circle (.1cm);
\draw[fill=black] (v2) circle (.07cm);
\end{tikzpicture}
\end{gathered}
\ = \ 
\begin{gathered}
\begin{tikzpicture}[line width=1.5 pt,node distance=1 cm and 1.5 cm]
\coordinate[label =left: $N$] (i1);
\coordinate[right= 1cm of i1](v1);
\coordinate[below= 0.5cm of v1,label= left:$ H$](vaux);
\coordinate[right = 1cm of v1, label= right:$ \ell_L$](f1);
\coordinate[below = 1 cm of v1](v2);
\coordinate[left = 1 cm of v2, label=left: $ Q_L$] (i2);
\coordinate[right =  1 cm of v2,label=right: $ t_R$] (f2);
\draw[fermionnoarrow] (i1) -- (v1);
\draw[scalar] (v2) -- (v1);
\draw[fermion] (v2) -- (i2);
\draw[fermion] (f2) -- (v2);
\draw[fermion] (f1) -- (v1);
\draw[fill=red] (v1) circle (.1cm);
\draw[fill=black] (v2) circle (.07cm);
\end{tikzpicture}
\end{gathered}
\ = \ 
\begin{gathered}
\begin{tikzpicture}[line width=1.5 pt,node distance=1 cm and 1.5 cm]
\coordinate[label =left: $N$] (i1);
\coordinate[right= 1cm of i1](v1);
\coordinate[below= 0.5cm of v1,label= left:$ H$](vaux);
\coordinate[right = 1cm of v1, label= right:$ \ell_L$](f1);
\coordinate[below = 1 cm of v1](v2);
\coordinate[left = 1 cm of v2, label=left: $ t_R$] (i2);
\coordinate[right =  1 cm of v2,label=right: $ Q_L$] (f2);
\draw[fermionnoarrow] (i1) -- (v1);
\draw[scalar] (v2) -- (v1);
\draw[fermion] (i2) -- (v2);
\draw[fermion] (v2) -- (f2);
\draw[fermion] (f1) -- (v1);
\draw[fill=red] (v1) circle (.1cm);
\draw[fill=black] (v2) circle (.07cm);
\end{tikzpicture}
\end{gathered}
\\
%%%%%%%%%%%%%%%%%%%%%%%%%%%%%%%%%%%
\gamma_{7}&:& \quad 
 \begin{gathered}
\begin{tikzpicture}[line width=1.5 pt,node distance=1 cm and 1.5 cm]
\coordinate[label =left: $\ell_L$] (i1);
\coordinate[below right= 1cm of i1](v1);
\coordinate[above right = 1cm of v1, label= right:$ H$](f1);
\coordinate[below left = 1 cm of v1, label=left: $ H$] (i2);
\coordinate[below right =  1 cm of v1,label=right: $ \ell_L$] (f2);
\draw[fermion] (i1) -- (v1);
\draw[scalar] (i2) -- (v1);
\draw[fermion] (f2) -- (v1);
\draw[scalar] (f1) -- (v1);
\draw[fill=gray] (v1) circle (.2cm);
\end{tikzpicture}
\end{gathered}  
=
\begin{gathered}
\begin{tikzpicture}[line width=1.5 pt,node distance=1 cm and 1.5 cm]
\coordinate[label =left: $\ell_L$] (i1);
\coordinate[below right= 1cm of i1](v1);
\coordinate[right = 0.5cm of v1,label=below:$N$](vaux);
\coordinate[right= 1 cm of v1](v2);
\coordinate[above right = 1cm of v2, label= right:$H$](f1);
\coordinate[below left = 1 cm of v1, label=left: $H$] (i2);
\coordinate[below right =  1 cm of v2,label=right: $ \ell_L$] (f2);
\draw[fermion] (i1) -- (v1);
\draw[fermionnoarrow] (v1) -- (v2);
\draw[scalar] (i2) -- (v1);
\draw[fermion] (f2) -- (v2);
\draw[scalar] (f1) -- (v2);
\draw[fill=red] (v1) circle (.1cm);
\draw[fill=red] (v2) circle (.1cm);
\end{tikzpicture}
\end{gathered}
+
\begin{gathered}
\begin{tikzpicture}[line width=1.5 pt,node distance=1 cm and 1.5 cm]
\coordinate[label =left: $\ell_L$] (i1);
\coordinate[right= 1cm of i1](v1);
\coordinate[below= 0.5cm of v1,label= left:$N$](vaux);
\coordinate[right = 1cm of v1, label= right:$H^\dagger$](f1);
\coordinate[below = 1 cm of v1](v2);
\coordinate[left = 1 cm of v2, label=left: $H$] (i2);
\coordinate[right =  1 cm of v2,label=right: $ \ell_L$] (f2);
\draw[fermion] (i1) -- (v1);
\draw[fermionnoarrow] (v1) -- (v2);
\draw[scalar] (i2) -- (v2);
\draw[fermion] (f2) -- (v2);
\draw[scalar] (f1) -- (v1);
\draw[fill=red] (v1) circle (.1cm);
\draw[fill=red] (v2) circle (.1cm);
\end{tikzpicture}
\end{gathered}  \quad = 
 \begin{gathered}
\begin{tikzpicture}[line width=1.5 pt,node distance=1 cm and 1.5 cm]
\coordinate[label =left:$\bar \ell_L$] (i1);
\coordinate[below right= 1cm of i1](v1);
\coordinate[above right = 1cm of v1, label= right:$ H$](f1);
\coordinate[below left = 1 cm of v1, label=left: $H^\dagger$] (i2);
\coordinate[below right =  1 cm of v1,label=right: $\ell_L$] (f2);
\draw[fermion] (v1) -- (i1);
\draw[scalar] (v1) -- (i2);
\draw[fermion] (v1) -- (f2);
\draw[scalar] (v1) -- (f1);
\draw[fill=gray] (v1) circle (.2cm);
\end{tikzpicture}
\end{gathered}  
\end{eqnarray*}

 \begin{eqnarray*}
%%%%%%%%%%%%%%%%%%%%%%%%%%%%%%%%%%%
\gamma_{8} &:& \quad
\begin{gathered}
\begin{tikzpicture}[line width=1.5 pt,node distance=1 cm and 1.5 cm]
\coordinate[label =left: $\ell_L$] (i1);
\coordinate[below right= 1cm of i1](v1);
\coordinate[above right = 1cm of v1, label= right:$ H$](f1);
\coordinate[below left = 1 cm of v1, label=left: $\ell_L$] (i2);
\coordinate[below right =  1 cm of v1,label=right: $H$] (f2);
\draw[fermion] (i1) -- (v1);
\draw[fermion] (i2) -- (v1);
\draw[scalar] (f2) -- (v1);
\draw[scalar] (f1) -- (v1);
\draw[fill=gray] (v1) circle (.2cm);
\end{tikzpicture}
\end{gathered}
\ \  = \ \
\begin{gathered}
\begin{tikzpicture}[line width=1.5 pt,node distance=1 cm and 1.5 cm]
\coordinate[label =left: $\ell_L$] (i1);
\coordinate[right= 1cm of i1](v1);
\coordinate[below= 0.5cm of v1,label= left:$N$](vaux);
\coordinate[right = 1cm of v1, label= right:$ H$](f1);
\coordinate[below = 1 cm of v1](v2);
\coordinate[left = 1 cm of v2, label=left: $\ell_L$] (i2);
\coordinate[right =  1 cm of v2,label=right: $H$] (f2);
\draw[fermion] (i1) -- (v1);
\draw[fermionnoarrow] (v1) -- (v2);
\draw[fermion] (i2) -- (v2);
\draw[scalar] (f2) -- (v2);
\draw[scalar] (f1) -- (v1);
\draw[fill=red] (v1) circle (.1cm);
\draw[fill=red] (v2) circle (.1cm);
\end{tikzpicture}
\end{gathered}
\quad   \! +
\begin{gathered}
\begin{tikzpicture}[line width=1.5 pt,node distance=1 cm and 1.5 cm]
\coordinate[label =left: $\ell_L$] (i1);
\coordinate[right= 1cm of i1](v1);
\coordinate[below= 0.5cm of v1,label= left:$N$](vaux);
\coordinate[right = 1cm of v1, label= right:$H$](f1);
\coordinate[below = 1 cm of v1](v2);
\coordinate[left = 1 cm of v2, label=left: $\ell_L$] (i2);
\coordinate[right =  1 cm of v2,label=right: $H$] (f2);
\draw[fermion] (i1) -- (v1);
\draw[fermionnoarrow] (v1) -- (v2);
\draw[fermion] (i2) -- (v2);
\draw[scalaruchannel2] (f2) -- (v1);
\draw[scalaruchannel2] (f1) -- (v2);
\draw[fill=red] (v1) circle (.1cm);
\draw[fill=red] (v2) circle (.1cm);
\end{tikzpicture}
\end{gathered} \quad  \  = \  
\begin{gathered}
\begin{tikzpicture}[line width=1.5 pt,node distance=1 cm and 1.5 cm]
\coordinate[label =left: $ \ell_L$] (i1);
\coordinate[below right= 1cm of i1](v1);
\coordinate[above right = 1cm of v1, label= right:$ H$](f1);
\coordinate[below left = 1 cm of v1, label=left: $ \ell_L$] (i2);
\coordinate[below right =  1 cm of v1,label=right: $H$] (f2);
\draw[fermion] (v1) -- (i1);
\draw[fermion] (v1) -- (i2);
\draw[scalar] (v1) -- (f2);
\draw[scalar] (v1) -- (f1);
\draw[fill=gray] (v1) circle (.2cm);
\end{tikzpicture}
\end{gathered}\\
\gamma_{9} &:& \quad
\begin{gathered}
\begin{tikzpicture}[line width=1.5 pt,node distance=1 cm and 1.5 cm]
\coordinate[label =left: $N$] (i1);
\coordinate[below right= 1cm of i1](v1);
\coordinate[above right = 1cm of v1, label= right:$\ell_L$](f1);
\coordinate[below left = 1 cm of v1, label=left: $H$] (i2);
\coordinate[below right =  1 cm of v1,label=right: $A_\mu$] (f2);
\draw[fermionnoarrow] (i1) -- (v1);
\draw[scalar] (v1) -- (i2);
\draw[vector] (f2) -- (v1);
\draw[fermion] (v1) -- (f1);
\draw[fill=gray] (v1) circle (.2cm);
\end{tikzpicture}
\end{gathered}
\ = \ 
\begin{gathered}
\begin{tikzpicture}[line width=1.5 pt,node distance=1 cm and 1.5 cm]
\coordinate[label =left: $N$] (i1);
\coordinate[below right= 1cm of i1](v1);
\coordinate[right= 0.5cm of v1,label=below:$\ell_L$](aux);
\coordinate[right= 1cm of v1](v2);
\coordinate[above right = 1cm of v2, label= right:$\ell_L$](f1);
\coordinate[below left = 1 cm of v1, label=left: $H$] (i2);
\coordinate[below right =  1 cm of v2,label=right: $A_\mu$] (f2);
\draw[fermionnoarrow] (i1) -- (v1);
\draw[scalar] (v1) -- (i2);
\draw[fermion](v1)--(v2);
\draw[vector] (f2) -- (v2);
\draw[fermion] (v2) -- (f1);
\draw[fill=red] (v1) circle (.1cm);
\draw[fill=cyan] (v2) circle (.1cm);
\end{tikzpicture}
\end{gathered}
\ + \ 
\begin{gathered}
\begin{tikzpicture}[line width=1.5 pt,node distance=1 cm and 1.5 cm]
\coordinate[label =left: $N$] (i1);
\coordinate[right= 1cm of i1](v1);
\coordinate[below=1cm of v1](v2);
\coordinate[below = 0.5cm of v1, label= left:$H$](aux);
\coordinate[right = 1cm of v1, label= right:$\ell_L$](f1);
\coordinate[left = 1 cm of v2, label=left: $H$] (i2);
\coordinate[right =  1 cm of v2,label=right: $A_\mu$] (f2);
\draw[fermionnoarrow] (i1) -- (v1);
\draw[scalar] (v2) -- (i2);
\draw[scalar] (v1)--(v2);
\draw[vector] (f2) -- (v2);
\draw[fermion] (v1) -- (f1);
\draw[fill=red] (v1) circle (.1cm);
\draw[fill=black] (v2) circle (.07cm);
\end{tikzpicture}
\end{gathered}\\
\gamma_{10} &:& \quad
\begin{gathered}
\begin{tikzpicture}[line width=1.5 pt,node distance=1 cm and 1.5 cm]
\coordinate[label =left: $N$] (i1);
\coordinate[below right= 1cm of i1](v1);
\coordinate[above right = 1cm of v1, label= right:$H$](f1);
\coordinate[below left = 1 cm of v1, label=left: $\ell_L$] (i2);
\coordinate[below right =  1 cm of v1,label=right: $A_\mu$] (f2);
\draw[fermionnoarrow] (i1) -- (v1);
\draw[fermion] (i2) -- (v1);
\draw[vector] (f2) -- (v1);
\draw[scalar] (f1) -- (v1);
\draw[fill=gray] (v1) circle (.2cm);
\end{tikzpicture}
\end{gathered}
\ \  = \ \
\begin{gathered}
\begin{tikzpicture}[line width=1.5 pt,node distance=1 cm and 1.5 cm]
\coordinate[label =left: $N$] (i1);
\coordinate[below right= 1cm of i1](v1);
\coordinate[right = 1cm of v1](v2);
\coordinate[right = 0.5cm of v1,label=below:$H$](aux);
\coordinate[above right = 1cm of v2, label= right:$H$](f1);
\coordinate[below left = 1 cm of v1, label=left: $\ell_L$] (i2);
\coordinate[below right =  1 cm of v2,label=right: $A_\mu$] (f2);
\draw[fermionnoarrow] (i1) -- (v1);
\draw[scalar] (v2)--(v1);
\draw[fermion] (i2) -- (v1);
\draw[vector] (f2) -- (v2);
\draw[scalar] (f1) -- (v2);
\draw[fill=red] (v1) circle (.1cm);
\draw[fill=black] (v2) circle (.07cm);
\end{tikzpicture}
\end{gathered}
\ + \ 
\begin{gathered}
\begin{tikzpicture}[line width=1.5 pt,node distance=1 cm and 1.5 cm]
\coordinate[label =left: $N$] (i1);
\coordinate[right= 1cm of i1](v1);
\coordinate[below = 1cm of v1](v2);
\coordinate[below=0.5cm of v1,label=left:$\ell_L$](aux);
\coordinate[right = 1cm of v1, label= right:$H$](f1);
\coordinate[left = 1 cm of v2, label=left: $\ell_L$] (i2);
\coordinate[right =  1 cm of v2,label=right: $A_\mu$] (f2);
\draw[fermionnoarrow] (i1) -- (v1);
\draw[fermion](v2)--(v1);
\draw[fermion] (i2) -- (v2);
\draw[vector] (f2) -- (v2);
\draw[scalar] (f1) -- (v1);
\draw[fill=red] (v1) circle (.1cm);
\draw[fill=cyan] (v2) circle (.1cm);
\end{tikzpicture}
\end{gathered}\\
\gamma_{11} &:& \quad
\begin{gathered}
\begin{tikzpicture}[line width=1.5 pt,node distance=1 cm and 1.5 cm]
\coordinate[label =left: $N$] (i1);
\coordinate[below right= 1cm of i1](v1);
\coordinate[above right = 1cm of v1, label= right:$H$](f1);
\coordinate[below left = 1 cm of v1, label=left: $A_\mu$] (i2);
\coordinate[below right =  1 cm of v1,label=right: $\ell_L$] (f2);
\draw[fermionnoarrow] (i1) -- (v1);
\draw[vector] (i2) -- (v1);
\draw[fermion] (v1) -- (f2);
\draw[scalar] (v1) -- (f1);
\draw[fill=gray] (v1) circle (.2cm);
\end{tikzpicture}
\end{gathered}
\ = \
\begin{gathered}
\begin{tikzpicture}[line width=1.5 pt,node distance=1 cm and 1.5 cm]
\coordinate[label =left: $N$] (i1);
\coordinate[right= 1cm of i1](v1);
\coordinate[below = 1 cm of v1](v2);
\coordinate[below=0.5cm of v1,label=left:$\ell_L$](aux);
\coordinate[right = 1cm of v1, label= right:$H$](f1);
\coordinate[left = 1 cm of v2, label=left: $A_\mu$] (i2);
\coordinate[right =  1 cm of v2,label=right: $\ell_L$] (f2);
\draw[fermionnoarrow] (i1) -- (v1);
\draw[fermion](v1)--(v2);
\draw[vector] (i2) -- (v2);
\draw[fermion] (v2) -- (f2);
\draw[scalar] (v1) -- (f1);
\draw[fill=red] (v1) circle (.1cm);
\draw[fill=cyan] (v2) circle (.1cm);
\end{tikzpicture}
\end{gathered}
\ + \
\begin{gathered}
\begin{tikzpicture}[line width=1.5 pt,node distance=1 cm and 1.5 cm]
\coordinate[label =left: $N$] (i1);
\coordinate[right= 1cm of i1](v1);
\coordinate[below=0.5cm of v1,label=left:$H$](aux);
\coordinate[below = 1 cm of v1](v2);
\coordinate[right = 1cm of v1, label= right:$H$](f1);
\coordinate[left = 1 cm of v2, label=left: $A_\mu$] (i2);
\coordinate[right =  1 cm of v2,label=right: $\ell_L$] (f2);
\draw[fermionnoarrow] (i1) -- (v1);
\draw[scalar](v1)--(v2);
\draw[vector] (i2) -- (v2);
\draw[fermionuchannel1] (v1) -- (f2);
\draw[scalaruchannel1] (v2) -- (f1);
\draw[fill=red] (v1) circle (.1cm);
\draw[fill=black] (v2) circle (.07cm);
\end{tikzpicture}
\end{gathered}
\end{eqnarray*}

%%%%%%%%%

%%%%%%%%%%%%%%%
%%%%%%%%%%%%%%%%%%%%%%%%%%%%%%%%%%%%%%%%%%%%%%%%%%%%%%%%%%%%%%%%%%%%%%%%%%%%%%%%
%%%%%%%%%%%%%%%%%%%%%%%%%%
% BIBLIOGRAPHY
\bibliography{Leptogenesis_U1L}
%%%%%%%%%%%%%%%%%%%%%%%%%%%%

\end{document}